\documentclass{article}                                                                           

\usepackage[margin=1.3in]{geometry}                                                     

\usepackage{setspace}                                                                             

\usepackage{indentfirst}                                                                          
\renewcommand\thesection{\Roman{section}}                                         
\renewcommand\thesubsection{\thesection.\Alph{subsection}}                   
\renewcommand\thesubsubsection{\thesubsection.\arabic{subsubsection}} 

\usepackage[explicit]{titlesec}                                                                  
\titleformat{\section}{\bfseries}{\thesection.}{1em}{\MakeUppercase{#1}}  
\titleformat{\subsection}{\bfseries}{\thesubsection.}{1em}{#1}                   
\titleformat{\subsubsection}{\itshape}{\thesubsubsection.}{1em}{#1}          

\usepackage{lastpage}                                                                             
\usepackage[figure,table]{totalcount}                                                        

\usepackage[]{footmisc}                                                                          

\usepackage{caption}                                                                              
\usepackage[labelformat=simple]{subcaption}                                           
\usepackage{standalone}
\usepackage{xifthen}     

\usepackage[english]{babel}	
\usepackage[utf8]{inputenc} 
\usepackage[kerning,spacing,babel]{microtype} 
\usepackage{eepic}      
\usepackage{epic}       
\usepackage[T1]{fontenc} 
\usepackage{lmodern}

\usepackage{authblk}		

\usepackage{upgreek}		
\usepackage[makeroom]{cancel}			

\usepackage{isotope}		
\usepackage[version=3]{mhchem}	

\usepackage{xspace}			
\usepackage{icomma}			
\usepackage{chngpage}		
\usepackage[usenames,dvipsnames,svgnames,table]{xcolor}			
\usepackage{enumitem}       
\usepackage{footnote}       

\usepackage[singlelinecheck=false, labelsep=quad]{caption}	
\usepackage{subcaption}		
\usepackage{placeins}   	
\usepackage{float}      	

\usepackage{supertabular}	
\usepackage{booktabs}		
\usepackage{array}			
\usepackage{multirow}		

\usepackage{graphicx}		
\usepackage{wrapfig}		
\usepackage{pgfplots}		
\usepackage{siunitx}

\usepackage{algorithmicx} 
\usepackage{listings} 	
\usepackage{verbatim}   

\usepackage{url}        
\usepackage{import}			
\usepackage{afterpage}  
\usepackage{lscape}     
\usepackage{rotating}   
\usepackage{chngcntr}   
\usepackage{appendix}   
\usepackage{titlesec}   

\usepackage{epsfig}     
\usepackage{epsf}       

\usepackage{hyperref}		

\usepackage{geometry}		
\usepackage{algorithm}  

\usepackage{amsmath}    
\usepackage{amsfonts}   
\usepackage{amssymb}    %
\usepackage{amsthm}     

\usepackage{nameref}		
\usepackage[capitalize,nameinlink]{cleveref}   

\makeatletter
\g@addto@macro\@floatboxreset\centering
\newcommand*{\currentname}{\@currentlabelname}
\makeatother

\newlength{\figheight}
\hypersetup{
  linktocpage=true,       
  bookmarks=true,         
  unicode=true,           
  pdftoolbar=true,        
  pdfmenubar=true,        
  pdffitwindow=false,     
  pdfstartview={FitH},    
  pdfnewwindow=true,      
  colorlinks=false,       
  linkcolor=red,          
  citecolor=green,        
  filecolor=magenta,      
  urlcolor=cyan,          
  pdfborder={0 0 0},      
}

\pgfplotsset{compat=1.14}

\lstdefinestyle{cpp}{
  belowcaptionskip=1\baselineskip,
  breaklines=true,
  frame=TB,
  xleftmargin=\parindent,
  language=C,
  showstringspaces=false,
  basicstyle=\footnotesize\ttfamily,
  keywordstyle=\bfseries\color{green!40!black},
  commentstyle=\itshape\color{purple!40!black},
  identifierstyle=\color{blue},
  stringstyle=\color{orange},
}

\lstdefinestyle{python}{
  belowcaptionskip=1\baselineskip,
  breaklines=true,
  frame=,
  xleftmargin=\parindent,
  language=Python,
  showstringspaces=false,
  basicstyle=\footnotesize\ttfamily,
  keywordstyle=\bfseries\color{blue},
  commentstyle=\itshape\color{gray},
  identifierstyle=\color{black},
  stringstyle=\color{green},
}

\lstdefinelanguage{Moose}{
  morekeywords={one,two,three,four,five,six,seven,eight,
  nine,ten,eleven,twelve,o,clock,rock,around,the,tonight},
  sensitive=false,
  morecomment=[l]{//},
  morecomment=[s]{/*}{*/},
  morestring=[b]",
}

\lstdefinestyle{moose}{
  belowcaptionskip=1\baselineskip,
  breaklines=true,
  frame=tb,
  xleftmargin=\parindent,
  language=Moose,
  showstringspaces=false,
  basicstyle=\footnotesize\ttfamily,
  keywordstyle=\bfseries\color{blue},
  commentstyle=\itshape\color{gray},
  identifierstyle=\color{black},
  stringstyle=\color{green},
}

\newcommand{\rattlesnake}{Rattlesnake\xspace}
\newcommand{\saaft}{SAAF\(\tau\)\xspace}

\newcommand{\parenthesis}[1]{{\left(#1 \right)}}

\newcommand{\grad}{\vec{\nabla}}
\newcommand{\del}{\vec{\nabla}}
\newcommand{\adj}[2][{}]{{{#2}^{\dagger#1}}}

\newcommand{\iter}[1]{^{#1}}

\newcommand{\abs}[1]{\left|#1\right|}
\newcommand{\norm}[2][{}]{\lVert#2\rVert_{#1}}

\newcommand{\tento}[1]{\ensuremath{10^{#1}}\xspace}

\newcommand{\half}[1][1]{\frac{#1}{2}}

\newcommand{\op}[1]{\ensuremath{\mathbf{\mathcal{#1}}}}

\newcommand{\dx}[1][x]{\,d#1}

\newcommand{\dmu}{\dx[\mu]}
\newcommand{\domg}{\dx[\direction]}

\newcommand{\intsp}{\int_{4\pi}}

\newcommand{\intpolar}{\int_{-1}^{1}}

\newcommand*{\domain}{\ensuremath{\mathcal{D}}\xspace}
\newcommand*{\boundary}{\ensuremath{{\partial\domain}}\xspace}

\newcommand{\testfct}{\ensuremath{\phi^{*}}\xspace}
\newcommand{\atestfct}{\ensuremath{\psi^{*}}\xspace}
\newcommand{\normal}{\ensuremath{\vec{n}}\xspace}

\newcommand{\sn}[1][N]{\ensuremath{S_#1}\xspace}

\newcommand{\keff}{\ensuremath{k_{\text{eff}}}\xspace}

\newcommand{\addindex}[1]{\ifthenelse{\isempty{#1}}{}{,{#1}}}
\newcommand{\spharmonic}{\ensuremath{\mathrm{Y}_{l}^{p}\xspace}}
\newcommand{\mflux}[1][]{\ensuremath{\phi_{l\addindex{#1}}^{p}\xspace}}
\newcommand{\sigl}[2][{}]{\ensuremath{\sigma_{#2\addindex{#1}}}\xspace}

\newcommand{\addgroup}[1]{\ifthenelse{\isempty{#1}}{}{_{#1}}}
\newcommand{\direction}{\ensuremath{\vec{\Omega}}\xspace}

\newcommand{\current}[1][]{\ensuremath{\vec{J}\addgroup{#1}}\xspace}

\newcommand{\drift}[1][]{\ensuremath{\hat{D}\addgroup{#1}}\xspace}
\newcommand{\DC}[1][]{\ensuremath{\mathrm{D}\addgroup{#1}}\xspace}

\newcommand{\xslabel}[2][]{\ifthenelse{\isempty{#1}}{\mathrm{#2}}{\mathrm{#2},#1}}
\newcommand{\sigt}[1][]{\ensuremath{\sigma_{\xslabel[#1]{t}}}\xspace}
\newcommand{\sigs}[1][]{\ensuremath{\sigma_{\xslabel[#1]{s}}}\xspace}
\newcommand{\sigf}[1][]{\ensuremath{\sigma_{\xslabel[#1]{f}}}\xspace}

\newcommand{\siga}[1][]{\ensuremath{\sigma_{\xslabel[#1]{a}}}\xspace}
\newcommand{\sigtr}[1][]{\ensuremath{\sigma_{\xslabel[#1]{tr}}}\xspace}

\newcommand{\weight}[1][]{\ensuremath{\mathrm{w}\addgroup{#1}}\xspace}

\newcommand{\unit}[1]{\ensuremath{\mathrm{#1}}\xspace}

\newcommand{\cm}{\,\unit{cm}}

\newcommand{\sfluxunit}{\,\ensuremath{\frac{1}{\unit{cm}^2\unit{s}}}}
\newcommand{\afluxunit}{\,\ensuremath{\frac{1}{\unit{cm}^2\unit{s}\cdot\unit{st}}}}

\newcommand{\Xsunit}{\,\ensuremath{\unit{\frac{1}{cm}}}}
\newcommand{\sourceunit}{\,\ensuremath{\unit{\frac{\mathrm{n}}{s}}}}

\hypersetup{
    pdftitle={A Weighted Least-Squares Transport Equation Compatible with Source Iteration and Voids},
    pdfauthor={Hans R Hammer},
    pdfsubject={Nonlinear diffusion acceleration for Least squares in voids},
    pdfkeywords={Void, Least Square, NDA},
}

\renewcommand{\drift}[1][]{\ensuremath{\hat{\alpha}\addgroup{#1}}\xspace}

\title{A Weighted Least-Squares Transport Equation Compatible with Source Iteration and Voids}
\author[1,2]{Hans R. Hammer\thanks{email: hrhammer@lanl.gov}\xspace}
\author[2]{Jim E. Morel\thanks{email: morel@tamu.edu}\xspace}
\author[3]{Yaqi Wang\thanks{email: yaqi.wang@inl.gov}\xspace}
\affil[1]{Los Alamos National Laboratory - T3: Fluid Dynamics and Solid Mechanics\\ Bikini Atoll Road, Los Alamos, NM, 87545}
\affil[2]{Texas A\&M University - Department of Nuclear Engineering\\ 3133 TAMU, College Station, TX 77843-3133}
\affil[3]{Idaho National Laboratory, 1955 N. Fremont Ave, Idaho Falls, ID 83415}
\date{                               
    \vspace{12cm}
}

\begin{document}
    
    \clearpage\maketitle
    \thispagestyle{empty}
    \pagebreak
    
    \begin{abstract}
        Second order forms of the transport equation allow the use of continuous finite elements (CFEM). This can be desired in multi-physics calculations where other physics require CFEM discretizations. Second-order transport operators are generally self-adjoint, yielding symmetric positive-definite matrices, which allow the use of efficient linear algebra solvers with an enormous advantage in memory usage. \par
        
        Least-squares (LS) forms of the transport equation can circumvent the void problems of other second order forms, but are almost always non-conservative. Additionally, the standard LS form is not compatible with discrete ordinates method (\sn) iterative solution techniques such as source iteration. A new form of the least-squares transport equation has recently been developed that is compatible with voids and standard \sn iterative solution techniques. Performing Nonlinear Diffusion Acceleration (NDA) using an independently-differenced low-order equation enforces conservation for the whole system, and makes this equation suitable for reactor physics calculations. In this context independent means that both the transport and low-order solutions converge to the same scalar flux and current as the spatial mesh is refined, but for a given mesh, the solutions are not necessarily equal.
        
        In this paper we show that introducing a weight function to this least-squares equation improves issues with causality and can render our equation equal to the Self-Adjoint Angular Flux (SAAF) equation.  Causality is a principle of the transport equation which states that information only travels downstream along characteristics.  This principle can be violated numerically. We show how to limit the weight function in voids and demonstrate the effect of this limit on the accuracy. Using the C5G7 benchmark, we compare our method to the self-adjoint angular flux formulation with a void treatment (\saaft), which is not self-adjoint and has a non-symmetric coefficient matrix. We show that the weighted least-squares equation with NDA gives acceptable accuracy relative to the \saaft equation while maintaining a symmetric positive-definite system matrix.
        
        \vspace{1em}\noindent\textbf{Keywords} --- Neutron Transport, Weighted Least-Squares, Nonlinear Diffusion Acceleration, Voids, HOLO
    \end{abstract}
    
    \pagebreak
    
    \section{Introduction}

    Second order forms of the transport equation offer a stable discretization using continuous finite elements (CFEM), which is especially appealing in the context of multi-physics calculations within frameworks with well-developed support for CFEM. One example is Rattlesnake~\cite{wang_hybrid_2017,wang_diffusion_2014}, Idaho National Laboratory's transport code within the MOOSE framework~\cite{gaston_moose:_2009}, which supports several second order schemes. Additionally, a second order form is generally compatible with discretizations that result in symmetric positive-definite (SPD) matrices~\cite{gesh_finite_1999} for the standard source iteration equations with \sn discretization. SPD matrices can then be solved by highly efficient linear algebra solvers, especially the conjugate gradient method~\cite{saad_iterative_2003} with preconditioning. The conjugate gradient algorithm only requires the storage of three solutions vectors, which is a large advantage compared with the general GMRES algorithm for non-SPD matrices and thus can also offer better convergence since no restart is necessary. \par
    
    Current developments in modeling and simulation raised the needs for tools which are able to handle voids or near voids. While this is definitely possible with the first order transport equation, second order schemes often show singularities and conditioning or convergence problems for very small (near zero) total cross sections~\cite{ackroyd_treatment_1986,morel_self-adjoint_1999}. Certain least-squares (LS) forms of the transport equation can circumvent the void problems of other second order forms, but are non-conservative, which explains why they are not commonly used in the nuclear community. Additionally, the left-hand side of the standard LS equations are coupled between all directions due to scattering, preventing the use of standard \sn iterative solution techniques. A newly developed form of the least-squares transport equation is compatible with voids and standard \sn iterative solution techniques, but is also non-conservative \cite{hansen_least-squares_2014}. Conservation of particles is only achieved as the numerical solution converges to the analytical solution. Conservation is of the utmost importance for criticality calculations, and if not enforced, can lead to large errors in the critical eigenvalue (\keff)~\cite{peterson_conservative_2015} and the flux. \par
    
    Source iteration is a common and well proven method to iteratively solve the discrete ordinates equations~\cite{adams_fast_2002}. Over the years researchers developed many improvements to this simple solution process. Many addressed the slow iterative convergence of the source iterations for highly diffusive media, for example, via diffusion synthetic acceleration (DSA) or nonlinear diffusion acceleration (NDA). In addition, the use of an inconsistent, but conservative NDA low-order equation enforces conservation for the whole system as shown by Peterson et al.~\cite{peterson_conservative_2015} and therefore is an important improvement for the weighted LS equation even in non-diffusive cases. The NDA method is especially of interest for reactor physics problems, since it is easily adopted to solve criticality problems ~\cite{park_nonlinear_2012}, and enforces conservation of particles for the WLS equation.  Linear DSA does not seem to be appropriate for criticality problems if one wants to perform the \keff-calculation with the low-order diffusion equations. Voids are problematic for both DSA and NDA because the standard diffusion coefficient is unbounded in voids. \par
    
    The purpose of this paper is to develop a weighted least-squares transport formulation that is compatible with voids. We have also developed a conservative void-compatible NDA scheme for our weighted least-squares transport formulation.  However, in this paper we focus upon our weighted least-squares transport formulation.  The void-compatible NDA scheme will be described in a separate paper \cite{hammer_nonlinear_2018}. Instead we use the conservative NDA scheme of Peterson et al.~\cite{peterson_conservative_2015} in this paper, which is not void-compatible.
    
    In this paper we show that introducing a weight function to this LS equation improves issues with causality and can render our equation equal to the Self-Adjoint Angular flux (SAAF) equation~\cite{morel_self-adjoint_1999}. Causality is a principle of the transport equation which states that information only travels downstream along the direction of neutron travel. This principle can be violated numerically. We show how to limit the weight function in voids and the effect of this limit on accuracy. Using the C5G7 benchmark we compare our method against the self-adjoint angular flux formulation with a void treatment (\saaft), which was introduced by Wang~\cite{wang_diffusion_2014}. This formulation uses a first order derivative for stabilization in optically thin cells, which results in a non-symmetric coefficient matrix. We demonstrate that the weighted least-squares equation with NDA gives acceptable accuracy relative to the \saaft equation while maintaining a symmetric positive-definite system matrix. \par
    
    In the next section we first derive the \saaft equation, and show that it can reduce to the first-order equation or the SAAF equation, depending upon the value of a certain parameter. Then we derive our weighted least-squares (WLS) equation, and show that it can reduce to the SAAF equation depending upon the definition of the weight function.  Our NDA scheme for the WLS equation is then derived.  Computational results are given in the next section.  These 
    results relate to following: a comparison of the weighted and unweighted least-squares equations; limiting of the weight function in voids; and a comparison of the \saaft and WLS methods applied to the C5G7 reactor physics benchmark.  In the final section we give conclusions and a summary of results.
    
    \section{Theory} \label{ch:second_order}

\subsection{Self-Adjoint Angular Flux Equation with void treatment} \label{sec:saaft}
    
    The standard SAAF equation \cite{morel_self-adjoint_1999} is not defined in voids. Wang et al.~\cite{wang_diffusion_2014} proposed a modified version of the SAAF equation that is well defined in voids. Here we shall give a short derivation of the self-adjoint angular-flux equation with void treatment (\saaft). Further details are described in the paper by Wang. This equation is used as comparison for our WLS equation. \par
    
    We will derive the steady-state mono-energetic \saaft equation for simplicity. The extension to multi-group is straightforward. Consider the first order transport equation in operator form, where \(\psi(\vec{x},\direction)\) is the angular flux with \(\vec{x} \in \domain\), \(\direction\in 4\pi\) (\(4\pi\) represents the entire 2D unit sphere)
    \begin{subequations} \label{eq:operator_form}
        \begin{equation} \label{eq:operator_transport}
        \op{L}\psi = \op{S}\psi + \op{F}\psi + \op{Q}
        \end{equation}
        where
        \begin{equation} \label{eq:operator_streaming}
        \op{L} \equiv \direction \cdot \del  + \sigt
        \end{equation}
        is the streaming and collision operator,
        \begin{equation}  \label{eq:operator_scattering}
        \op{S} \equiv \sum_{l=0}^{\infty}\sum_{p=-l}^{l} \frac{2l+1}{4\pi} \spharmonic\left(\direction\right) \sigl{l} \op{M}
        \end{equation}
        the scattering operator with
        \begin{equation}
        \op{M} \equiv \intsp \domg \,\spharmonic\left(\direction\right)
        \end{equation}
        the flux moments and the scattering moments
        \begin{equation}
        \sigl{l}\ \equiv 2\pi \intpolar \sigs\left(\mu\right) P_{l}\left(\mu\right) \dmu.
        \end{equation}
        Finally
        \begin{equation} \label{eq:operator_fission}
        \op{F} \equiv  \frac{1}{4\pi} \intsp \dx[\direction']\, \bar{\nu}\sigf
        \end{equation}
    \end{subequations}
    is the fission source operator and the distributed source is denoted with \op{Q}. Here, \sigt is the total cross section, \sigs is the scattering cross section, \sigf the fission cross section with the fission spectrum \(\chi\) and the average number of released neutrons \(\bar{\nu}\), \(\spharmonic\) are the spherical harmonics and \(P_{l}\) the Legendre polynomials.
    
    \Cref{eq:operator_transport} is then formally solved for the angular flux as follows
    \begin{equation} \label{eq:saaf_afe}
        \psi = -\frac{1}{\sigt}\direction \cdot \del \psi + \frac{1}{\sigt}\op{S}\psi + \frac{1}{\sigt}\op{F}\psi + \frac{1}{\sigt}\op{Q}.
    \end{equation}

    Next we define the stabilization parameter \(\tau\) as a function of a cell's optical thickness
    \begin{align} \label{eq:saaf_tau}
        \tau \equiv \begin{cases}
            \frac{1}{\sigt}, & \sigt h \ge \zeta \\
            \frac{h}{\zeta}, & \sigt h < \zeta
        \end{cases}
    \end{align}
    with \(\zeta\) the stabilization threshold, normally set to \(0.5\) as described in the reference~~\cite{wang_diffusion_2014}. We subtract and add \(\tau\sigt\psi\) to \(\psi\) to obtain
    \begin{equation} \label{eq:saaf_tmp}
        \psi = \left(1 - \tau\sigt\right)\psi + \tau\sigt\psi.
    \end{equation}
    Then we substitute \cref{eq:saaf_afe} into the last term of the \cref{eq:saaf_tmp} to obtain
    \begin{align} \label{eq:saaf_supg}
        \psi &= \left(1 - \tau\sigt\right)\psi + \tau\left(- \direction \cdot \del \psi + \op{S}\psi + \op{F}\psi + \op{Q} \right).
    \end{align}
    Substituting from \cref{eq:saaf_supg} into the streaming term of the transport equation we obtain the \saaft equation for one energy group
    \begin{multline} \label{eq:saaft_mono}
        -\direction \cdot \del \left[\tau\direction \cdot \del \psi\right]
        + \direction \cdot \del \left[\left(1 - \sigt\tau\right) \psi\right]
        + \sigt \psi
        = \sum_{l=0}^{L}\sum_{p=-l}^{l} \frac{2l+1}{4\pi} \spharmonic\left(\direction_m\right) \sigl{l} \mflux \\
        + \frac{1}{4\pi} \bar{\nu}\sigf\phi + \frac{q}{4\pi}
        - \direction \cdot \del \left[\tau \sum_{l=0}^{L}\sum_{p=-l}^{l} \frac{2l+1}{4\pi} \spharmonic\left(\direction\right) \sigl{l}  \mflux
        + \tau \frac{\bar{\nu}\sigf}{4\pi} \phi + \tau\frac{q}{4\pi}\right].
    \end{multline}
    
    This equation is compatible with voids, however the system matrix is not symmetric anymore due to the first derivative. It will reduce to the first order transport equation in the case of \(\tau = 0\). For \(\zeta = 0\) the stabilization parameter \cref{eq:saaf_tau} is \(\tau = \frac{1}{\sigt}\) and \cref{eq:saaft_mono} is equivalent to the standard SAAF form 
    \begin{multline} \label{eq:saaf_mono}
        -\direction \cdot \del \left[\frac{1}{\sigt}\direction \cdot \del \psi\right]
        + \sigt \psi
        = \sum_{l=0}^{L}\sum_{p=-l}^{l} \frac{2l+1}{4\pi} \spharmonic\left(\direction_m\right) \sigl{l} \mflux \\
        + \frac{1}{4\pi} \bar{\nu}\sigf\phi + \frac{q}{4\pi}
        - \direction \cdot \del \left[\frac{1}{\sigt} \sum_{l=0}^{L}\sum_{p=-l}^{l} \frac{2l+1}{4\pi} \spharmonic\left(\direction\right) \sigl{l}  \mflux
        + \frac{1}{\sigt} \frac{\bar{\nu}\sigf}{4\pi} \phi + \frac{1}{\sigt}\frac{q}{4\pi}\right].
    \end{multline}

    The weak form of the equation better defines the underlying physics than the classic partial-differential equation and can be converted to algebraic equations thus solved numerically with finite-dimensional function spaces. To derive the weak form we first multiply \cref{eq:saaft_mono} by a test function \(\atestfct\) and integrate over the whole domain. Using integration by parts on all terms containing a derivative we obtain:
    find \(\psi \in W_\domain\) such that
    \begin{multline} \label{eq:saaft_weak}
        \left(\tau\direction \cdot \del \psi, \direction \cdot \del \atestfct\right)_\domain
        + \left(\left(1 - \sigt\tau\right) \psi, \direction \cdot \del \atestfct\right)_\domain
        + \big(\sigt \psi,\atestfct \big)_\domain
        + \left<\psi, \left(\direction \cdot \normal\right) \atestfct \right>_{\boundary^+} \\
        = \left(\sum_{l=0}^{L}\sum_{p=-l}^{l} \frac{2l+1}{4\pi} \spharmonic\left(\direction\right) \sigl{l}  \mflux, \tau\direction \cdot \del \atestfct + \atestfct\right)_\domain
        + \left(\frac{\bar{\nu}\sigf}{4\pi}\phi, \tau\direction \cdot \del \atestfct + \atestfct\right)_\domain \\
        + \left(\frac{q}{4\pi}, \tau\direction \cdot \del \atestfct + \atestfct \right)_\domain
        - \left<\psi^{\mathrm{inc}}, \left(\direction \cdot \normal\right) \atestfct \right>_{\boundary^-},
    \end{multline}
    where the operator
    \begin{equation}
        \left(\vphantom{\del}\cdot, \cdot\right)_\domain \equiv \int_\domain \dx[V]
    \end{equation} 
    is the standard spatial inner product and 
    \begin{equation}
        \left<\vphantom{\del}\cdot, \cdot\right>_\boundary \equiv \oint_\boundary \dx[A]
    \end{equation} 
    is the corresponding surface integral over the boundary \(\boundary\). We further denote the incoming and outgoing boundary as
    \begin{subequations}
    \begin{align}
        \boundary^- &= \left\{\boundary \,\big|\, \direction \cdot \normal < 0 \right\} \\
        \boundary^+ &= \left\{\boundary \,\big|\, \direction \cdot \normal > 0 \right\}.
    \end{align}
\end{subequations}

\subsection{Weighted Least-Squares Method} \label{sec:wls}

    The standard least-squares (LS) form of the transport equation~\cite{drumm_least-squares_2011}
    \begin{equation} \label{eq:standard_ls}
        \adj{\left(\op{L} - \op{S}\right)}\left(\op{L} - \op{S}\right)\psi = \adj{\left(\op{L} - \op{S}\right)}q
    \end{equation}
    is not compatible with source iterations since the left hand side of the equation remains coupled in all directions. Here \(\adj{\left(\op{L} - \op{S}\right)}\) denotes the adjoint of the transport operator. The Least-Squares Equation derived by Hansen et al.~\cite{hansen_least-squares_2014} is a second order transport equation that is compatible with voids. In contrast to traditional least-squares forms this equation is also usable with source iterations with or without acceleration. \par
    
    Consider the first order transport equation in operator form as shown in \cref{eq:operator_form}. Under the standard inner product
    \begin{equation} \label{eq:inner_product}
        \left(\vphantom{\del}\cdot,\:\cdot\right) \equiv \int_\domain \int_{4\pi} \int_0^\infty~dE\, d\Omega\, dV.
    \end{equation}
    the adjoint of the streaming and collision operator \cref{eq:operator_streaming} is
    \begin{equation} \label{eq:op_adj}
        \adj{\op{L}} \equiv -\direction \cdot \del + \sigt.
    \end{equation}

    Multiplying \cref{eq:operator_transport} with a weight function \(\op{W}\) and the adjoint operator \cref{eq:op_adj} gives the weighted least-squares equation compatible with source iteration
    \begin{equation}
        \adj{\op{L}} \op{W} \op{L}\psi = \adj{\op{L}} \op{W} \op{S}\psi + \adj{\op{L}} \op{W} \op{F} \psi + \adj{\op{L}} \op{W}  \op{Q}.
    \end{equation} 
    Note that we only use the adjoint of the streaming and collision operator, and not the full adjoint to the transport equation. This gives us the ability to use source iterations and is the main difference relative to standard least-squares methods in \cref{eq:standard_ls}. The left-hand side of this equation is self-adjoint and decouples for all directions, if the scattering and fission source are lagged with Source Iterations. 
    The mono-energetic WLS equation can be written as
    \begin{subequations} \label{eq:wls_equation}
        \begin{multline} \label{eq:wls_transport}
            - \direction \cdot \del \left[ \weight \direction \cdot \del \psi\right]
            - \direction \cdot \psi \del \left[ \weight \sigt\right]
            + \weight \sigt^2\psi \\
            = -\direction\cdot\del\left[\weight \sum_{l=0}^{L}\sum_{p=-l}^{l} \frac{2l+1}{4\pi} \spharmonic\left(\direction\right) \sigl{l} \mflux
            + \weight\frac{\bar{\nu}\sigf}{4\pi} \phi
            + \weight\frac{q}{4\pi}\right] \\
            + \weight\sigt\sum_{l=0}^{L}\sum_{p=-l}^{l} \frac{2l+1}{4\pi} \spharmonic\left(\direction\right) \sigl{l} \mflux
            + \weight\frac{\sigt \bar{\nu}\sigf}{4\pi}\phi
            + \weight\frac{q\sigt}{4\pi}.
        \end{multline}
        where \weight denotes a weight function. The corresponding boundary conditions are
        \begin{align} \label{eq:wls_strong_bc}
            \psi\left(\vec{x}_b, \direction\right) &= \psi^\mathrm{inc}\left(\vec{x}_b, \direction \right), \qquad \forall\vec{x}_b \in \boundary, \quad \direction \cdot \normal < 0  \\
            \direction \cdot \del \psi\left(\vec{x}_b\right) + \sigt \psi\left(\vec{x}_b\right) &= \op{S}\psi\left(\vec{x}_b\right) + \op{F}\psi\left(\vec{x}_b\right) + \op{Q}\psi\left(\vec{x}_b\right),\qquad \direction \cdot \normal > 0.
        \end{align}
    \end{subequations}

    A rigorous property of transport solutions is causality. The definition of causality in this context is that the angular flux solution in a given direction is only influenced by upstream information. Given a point \(\vec{x}\) and a direction, \direction, the associated "upstream" points in an infinite medium are defined by 
    \begin{subequations}
        \begin{equation}
        \vec{x}_\mathrm{upstream} = \vec{x} - s\direction \qquad \forall s \in \mathcal{R}^+,
        \end{equation} 
        downstream is then accordingly
        \begin{equation}
            \vec{x}_\mathrm{downstream} = \vec{x} + s\direction \qquad \forall s \in \mathcal{R}^+.
        \end{equation} 
    \end{subequations}
    Second order forms however can be influenced by downstream information due to numerical errors. Problems with causality of these  formulations occur at material interfaces separating optically thin and optically thick material regions. This is a coarse mesh problem that decreases with increasing refinement of the mesh. The introduction of a weight function reduces this problem significantly~\cite{zheng_least-squares_2016}. Therefore we will use the weighted least-squares (WLS) equation with a weight function compatible with voids. The following weight function
    \begin{align} \label{eq:def_weight}
        \weight \equiv \frac{1}{\sigt}
    \end{align}
    improves the causality and makes our equations equivalent to the SAAF equation (\cref{eq:saaf_mono}). Using \cref{eq:wls_transport} with the weight function \cref{eq:def_weight} the first derivative term becomes
    \begin{equation}
        \direction \cdot \psi \del \left[ \weight \sigt\right] = \direction \cdot \psi \del \left[ 1\right] \equiv 0
    \end{equation}
    and the equation is equal to \cref{eq:saaf_mono}. However, the weight function \cref{eq:def_weight} is not defined in voids. To compensate for this we redefine the weight function to
    \begin{equation} \label{eq:def_void_weight}
        \weight \equiv \min\left(\frac{1}{\sigt},\,\weight[_\mathrm{max}]\right),
    \end{equation}
    where \(\weight_\mathrm{max} \) denotes a maximum value for the weight function. This definition will make the WLS equation well defined in voids and maintain the symmetric positive-definite properties of the resulting discretized matrix, that has to be inverted in the source iteration process. Therefore, the WLS equation is the same as the SAAF equation only for sufficiently  large total cross section \sigt.

    We derive the weak form by multiplying \cref{eq:wls_equation} with a test function \(\atestfct\) and integrate over the spatial domain \domain, applying integration by parts to all terms containing a derivative. Given a trial space \(W_\domain\), consisting of continuous basis functions, the weak form is as follows:
    Find \(\atestfct\in W_\domain\) such that
    \begin{multline} \label{eq:ls_weak_form_base}
        \left(\weight\direction\cdot\del\psi, \direction\cdot\del\atestfct
        + \sigt\atestfct\right)_\domain
        +\left(\weight\sigt\psi, \direction\cdot\del\atestfct + \sigt\atestfct\right)_\domain \\
        = \left(\weight\sum_{l=0}^{L}\sum_{p=-l}^{l}\frac{2l+1}{4\pi}\sigl{l} \mflux \spharmonic\left(\direction\right), \direction\cdot\del\atestfct + \sigt\atestfct\right)_\domain
        +\left(\weight\frac{\nu\sigf}{4\pi}\phi + \weight\frac{q}{4\pi}, \direction\cdot\del\atestfct + \sigt\atestfct\right)_\domain \\
        +\left<\direction\cdot\del\psi + \sigt\psi, \weight\left(\direction\cdot\normal\right)\atestfct \right>_{\boundary} \\
        -\left<\sum_{l=0}^{L}\sum_{p=-l}^{l}\frac{2l+1}{4\pi}\sigl{l} \spharmonic\left(\direction\right) \mflux
        + \frac{\nu\sigf}{4\pi} \phi + \frac{q}{4\pi}, \weight\left(\direction \cdot \normal\right) \atestfct\right>_{\boundary}.
    \end{multline}
    With the assumption that the first-order \(S_N\) transport equation is exactly satisfied on the boundary \boundary, all of the boundary terms cancel, leading to the following:
    \begin{multline}\label{eq:wls_weak}
        \left(\weight\direction\cdot\del\psi, \direction\cdot\del\atestfct + \sigt\atestfct\right)_\domain
         + \left(\weight\sigt\psi, \direction\cdot\del\atestfct + \sigt\atestfct\right)_\domain \\
        = \left(\weight\sum_{l=0}^{L}\sum_{p=-l}^{l}\frac{2l+1}{4\pi} \spharmonic\left(\direction\right) \sigl{l} \mflux, \direction\cdot\del\atestfct + \sigt\atestfct\right)_\domain \\
         + \left(\weight\frac{\nu\sigf}{4\pi}\phi + \weight\frac{q}{4\pi},  \direction\cdot\del\atestfct + \sigt\atestfct\right)_\domain
    \end{multline}
    An additional motivation for making this assumption is that it renders our Galerkin method for the second-order least-squares equation equivalent to the least-squares finite-element method for the first-order form of the \(S_N\) equations using the same trial space. \par

    The natural boundary condition of \cref{eq:wls_weak} is a Dirichlet boundary condition. This is difficult to implement in numerical codes, since it is ambiguous at boundary corners and edges. We chose to use the optional weak boundary condition
    \begin{equation} \label{eq:ls_weak_form_bc}
        \left<\weight f\left(\psi-\psi^{inc}\right),\psi^{*} \right>_{\boundary^-}
    \end{equation}
    instead, where \({\boundary^-}\) is the portion of the boundary for which \(\direction \cdot \normal < 0\). \par
    
    We define
    \begin{equation} \label{eq:ls_bc_1}
        f \equiv \sigt\left| \direction \cdot \vec{n}\right|
    \end{equation}
    based on the SAAF boundary condition~\cite{laboure_least-squares_2016}. However, the two SAAF boundary conditions are defined over the incoming and outgoing boundary respectively, while the optional WLS boundary conditions \cref{eq:ls_weak_form_bc} are only defined on the incoming boundary. Nevertheless, with this boundary factor the WLS equation with optional boundary condition is still equivalent to the SAAF equation. To demonstrate this, we add the boundary term \cref{eq:ls_weak_form_bc} to the left hand side of \cref{eq:wls_weak}, insert the weight function given by \cref{eq:def_weight}, and integrate \(\left(\vec{\Omega}\cdot \grad\psi,\psi^*\right)\) by parts to obtain
    \begin{multline} \label{eq:wls_saaft_compare}
        \left(\frac{1}{\sigt}\direction\cdot\del\psi, \direction\cdot\del\atestfct\right)_\domain
        + \big(\sigt\psi, \atestfct\big)_\domain \\
        + \left<\psi, \parenthesis{\direction \cdot \normal} \atestfct \right>_\boundary
        + \left<\left(\psi-\psi^{inc}\right),\abs{\direction\cdot\normal}\atestfct \right>_{\boundary^-}
        = 0.
    \end{multline}
    For simplification the source terms were set to zero. The boundary terms in \cref{eq:wls_saaft_compare} can now be manipulated as follows
    \begin{align}
        \left<\psi, \parenthesis{\direction \cdot \normal} \atestfct \right>_\boundary
        &+ \left<\left(\psi-\psi^{inc}\right),\abs{\direction\cdot\normal}\atestfct \right>_{\boundary^-} \notag \\
        &= \left<\psi, \parenthesis{\direction \cdot \normal} \atestfct \right>_{\boundary^+}
        + \left<\psi, \parenthesis{\direction \cdot \normal} \atestfct \right>_{\boundary^-} \notag\\
        &\qquad- \left<\psi, \parenthesis{\direction \cdot \normal} \atestfct \right>_{\boundary^-}
        + \left<\psi^{inc}, \parenthesis{\direction \cdot \normal} \atestfct \right>_{\boundary^-} \notag\\
        &= \left<\psi, \parenthesis{\direction \cdot \normal} \atestfct \right>_{\boundary^+}
        + \left<\psi^{inc}, \parenthesis{\direction \cdot \normal} \atestfct \right>_{\boundary^-}.
    \end{align}
    This is the SAAF boundary condition as stated in \cref{eq:saaft_weak}. \par
    
    For near void problems
    \begin{equation} \label{eq:ls_bc_7}
        f \equiv \max\left(\sigt, \frac{1}{h}\right)\left| \direction \cdot \vec{n}\right|
    \end{equation}
    gives a more accurate and better conditioned version. Here \(h\) denotes a characteristic length constant of the boundary cell, we used the maximum distance between cell vertices, however this might not be a good choice for cells with a large aspect ratio. Thus even with optional boundary condition the WLS equation is equivalent to the SAAF equation, if and only if all \sigt are larger than the thresholds in the weight and boundary functions. \par

    The resulting mono-energetic WLS equation used in this paper is defined as follows:
    Given a trial space \(W_\domain\), consisting of continuous basis functions, the weak form for a specific direction \(m = 1\dots M\) is as follows:
    Find \(\atestfct\in W_\domain\) such that
    \begin{multline}\label{eq:wls_weak_void}
        \left(\weight\direction\cdot\del\psi, \direction\cdot\del\atestfct + \sigt\atestfct\right)_\domain
        + \left(\weight\sigt\psi, \direction\cdot\del\atestfct + \sigt\atestfct\right)_\domain \\
        + \left<\weight f \left(\psi - \psi^\mathrm{inc}\right), \atestfct\right>_{\boundary^-}
        = \left(\weight\sum_{l=0}^{L}\sum_{p=-l}^{l}\frac{2l+1}{4\pi} \spharmonic\left(\direction\right) \sigl{l} \mflux, \direction\cdot\del\atestfct + \sigt\atestfct\right)_\domain \\
        + \left(\weight\frac{\nu\sigf}{4\pi}\phi,  \direction\cdot\del\atestfct + \sigt\atestfct\right)_\domain
        + \left(\weight\frac{q}{4\pi}, \direction\cdot\del\atestfct + \sigt\atestfct\right)_\domain
    \end{multline}
    with the weight function as described in \cref{eq:def_void_weight} and the boundary functions as described in \cref{eq:ls_bc_7}. Note that for \(\weight = 1\) the WLS equation reduces to the unweighted LS scheme. \par

	\subsection{Nonlinear Diffusion Acceleration}
	We derive our low order diffusion equation from the first order transport equation as shown by Peterson \cite{peterson_conservative_2015}. This results in an independently differenced, but conservative form of the NDA, which enforces conservation for the whole system. Integrating the mono-energetic transport equation over all angles gives us the zeroth moment equation
	\begin{equation} \label{eq:zeroth_moment}
		\del \cdot \current + \sigt \phi = \sigs \phi + q.
	\end{equation}
	To close \cref{eq:zeroth_moment}, we consider the first moment equation
	\begin{equation} \label{eq:first_moment}
		\sum_{m=1}^{M}\omega_m\direction_m\ \left(\direction_m\cdot \del \psi_{m, g}\right) + \sigt \current
		= \sigl{1}\current
	\end{equation}
	which gives the current
	\begin{equation} \label{eq:nda_current}
		\current = - \frac{1}{\sigtr} \sum_{m=1}^{M}\omega_m \direction_m \left(\direction_m \cdot \del \psi \right) 
	\end{equation}
	with the transport cross section
	\begin{equation} \label{eq:transport_xs}
		\sigtr \equiv \sigt - \sigl{1}.
	\end{equation}
	We use \cref{eq:nda_current} to construct an additive correction to Fick's law by adding and subtracting \(\DC \del \phi\)
	\begin{align} \label{eq:drift_current}
		\current =& -\DC \del \phi + \DC \del \phi
		-\frac{1}{\sigtr} \sum_{m=1}^{M} \omega_m \direction_m \direction_m \cdot \del \psi \notag \\
		=& -\DC \del \phi - \drift \phi
	\end{align}
	with the drift vector
	\begin{equation} \label{eq:drift_vector}
		\drift \equiv \frac{1}{\phi} \left(\frac{1}{\sigtr} \sum_{m=1}^{M}\omega_m\direction_m \left(\direction_m \cdot \del \psi \right) - \DC \del \phi \right)
	\end{equation}
	and the diffusion coefficient defined as
	\begin{equation} \label{eq:diffusion_coefficient}
		\DC \equiv \frac{1}{3\sigtr}.
	\end{equation}
	Substituting \cref{eq:drift_current} into \cref{eq:zeroth_moment} gives the NDA drift-diffusion equation
	\begin{equation} \label{eq:drift_diffusion}
		- \del \cdot \left[\DC \del \phi\right] - \del \cdot \left[\drift \phi\right] + \siga \phi = q.
	\end{equation}
	
	Multiplying \cref{eq:zeroth_moment} by a test function \(\phi^*\) and integrating over the domain gives the corresponding weak form
	\begin{equation} \label{eq:diffusion_weak_form}
		\left(\del \cdot \current, \testfct \right)_\domain + \left(\siga \phi, \testfct \right)_\domain 
		= \left(q, \testfct \right)_\domain.
	\end{equation}
	Applying integration by parts on the current term and substituting \cref{eq:drift_current} gives
	\begin{equation} \label{eq:low_order_eq}
		- \left(\DC \del \phi, \del\testfct \right)_\domain
		-\left(\drift \phi, \del\testfct \right)_\domain 
		+ \left< \normal \cdot \current, \testfct\right>_\boundary
		+ \left(\siga \phi, \testfct \right)_\domain
		=\left(q, \testfct \right)_\domain.
	\end{equation}\par
	
	The boundary term \(\left< \vec{n} \cdot \current, \phi^*\right>_\boundary\) still needs to be evaluated. While the reflective boundary condition is natural to the diffusion equation, the vacuum condition is more challenging. Using the partial currents we can define
	\begin{align} \label{eq:nda_bc}
		\left< \normal \cdot \current, \testfct \right>_{\boundary} 
		&= \left<J^\mathrm{\,out} - J^\mathrm{\,in}, \testfct \right>_{\boundary} \notag\\
		&= \left< \frac{1}{4}\kappa \phi - J^\mathrm{\,in}, \testfct \right>_{\boundary}
	\end{align}
	with the vacuum boundary coefficient as
	\begin{align} \label{eq:kappa}
		\kappa &\equiv 4\frac{J^\mathrm{\,out}}{\phi} \notag\\
		&= \frac{4}{\phi} \sum_{\normal \cdot \direction_m > 0} \omega_m \left(\normal \cdot \direction_m\right) \psi
	\end{align}
	and substitute this into the boundary term. Note that we changed the vacuum boundary coefficient from the one described in \cite{peterson_conservative_2015} to \cref{eq:kappa} to be consistent with the \saaft implementation in Rattlesnake~\cite{wang_diffusion_2014}.\par
	
	For a given iteration \(k\) the NDA scheme is defined as follows:
	\begin{enumerate}
		\item Solve the WLS transport equation
		\begin{subequations} \label{eq:nda}
			\begin{multline} \label{eq:nda_wls}
				\left(\weight\direction\cdot\del\psi\iter{k+\half} + \weight\sigt\psi\iter{k+\half}, \direction\cdot\del\atestfct + \sigt\atestfct\right)_\domain \\
				+\left<\weight f\left(\psi\iter{k+\half} - \psi^\mathrm{inc}\right),\atestfct \right>_{\boundary^-} \\
				= \left(\weight\sum_{l=0}^{L}\sum_{p=-l}^{l}\frac{2l+1}{4\pi} \spharmonic \sigl{l} \phi_{l}^{p,k}, \direction\cdot\del\atestfct + \sigt\atestfct\right)_\domain \\
				+ \left(\weight\frac{1}{4\pi}\nu\sigf\phi\iter{k} + \weight\frac{1}{4\pi}q,  \direction\cdot\del\atestfct + \sigt\atestfct\right)_\domain,\, m=1\dots M
			\end{multline}
			\item Calculate the correction terms for the diffusion equation
			\begin{equation} \label{eq:nda_kappa}
				\kappa\iter{k+\half} = \frac{4}{\phi} \sum_{\normal \cdot \direction_m > 0} \omega_m \left(\normal \cdot \direction_m\right) \psi\iter{k+\half}
			\end{equation}
			\begin{equation} \label{eq:nda_drift}
				\drift\iter{k+\half} = \frac{1}{\phi\iter{k+\half}} \left(\frac{1}{\sigtr} \sum_{m=1}^{M}\omega_m\direction_m \left(\direction_m \cdot \del \psi\iter{k+\half}_{m} \right) - \DC \del \phi\iter{k+\half} \right)
			\end{equation}
			\item Solve the diffusion equation
			\begin{multline} \label{eq:nda_diffusion}
				- \left(\DC \del \phi\iter{k+1}, \del\testfct \right)_\domain
				-\left(\drift\iter{k+\half} \phi\iter{k+1}, \del\testfct \right)_\domain  \\
				+ \left< \frac{\kappa\iter{k+\half}}{4}\phi\iter{k+1} - J^\mathrm{in}, \testfct\right>_\boundary
				+ \left(\siga \phi\iter{k+1}, \testfct \right)_\domain
				= \left(q, \testfct \right)_\domain
			\end{multline}
			\item Check convergence and update the scattering source
            \begin{equation}
                \frac{\norm{\phi\iter{l+1} - \phi\iter{l}}}{\phi\iter{l}} \le \tau
            \end{equation}
		\end{subequations}
	\end{enumerate}
	
	The iteration scheme for the NDA starts with a low order solve of \cref{eq:nda_diffusion} assuming \(\drift\iter{\half} = 0\) and \(\kappa\iter{\half} = 1\). The scalar flux is transferred to the high order system and used for the scattering and fission source. The new angular flux is obtained and the drift vector and boundary coefficient calculated. These are then used for the next low order diffusion solve. This iteration continues until convergence of the low order and high order solutions.\par
	
	The derivation of the multi-group equations is similar. The only thing to be considered are the crossgroup scattering terms in the drift vector
	\begin{equation} \label{eq:drift_vector_mg}
		\drift[g]\iter{k+\half} \equiv \frac{1}{\phi_{g}\iter{k+\half}} \left(\frac{1}{\sigtr[g]} \sum_{m=1}^{M}\omega\direction\left(\direction\cdot \del \psi_{m, g}\iter{k+\half}\right)
		-\frac{1}{\sigtr[g]}\sum_{\substack{g'=1\\g' \neq g}}^{G} \sigma_{1, g' \rightarrow g} \current_{g'}\iter{k+\half} 
		- \DC[g] \del\phi_{g}\iter{k+\half}\right).
	\end{equation}
    \section{Numerical Results} \label{ch:results}

\subsection{Implementation in \rattlesnake}
    The high-order low-order system is represented by two sets of equation systems in Rattlesnake. The low-order diffusion equation is solved with the PJFNK solver~\cite{park_nonlinear_2012} preconditioned with Hypre BoomerAMG~\cite{henson_boomeramg:_2000}. BoomerAMG is not  specifically designed for the non-symmetric Jacobian of this system. Currently there is a study in progress to replace it with an AMG solver for anisotropic diffusion~\cite{de_sterck_smoothed_2010,manteuffel_root-node_2016}, which is better suited for non-axis aligned problems. Nevertheless, for now BoomerAMG is an effective preconditioner. Before the actual solve \rattlesnake performs several free power iterations. The power iterations ensure that the initial guess for the Newton solve is close to the largest eigenvalue. All consecutive solves use Picard iterations. Each Picard iterations consists of the steps corresponding to \cref{eq:nda}. First the low-order equation is solved using a nonlinear eigenvalue solver~\cite{knoll_acceleration_2011} to obtain an initial guess for the scalar flux. The scalar flux is transfered to the high-order system. The left hand side of the transport equation, the streaming and collision part, is inverted. Finally, the drift vector and boundary coefficients are updated using the angular fluxes. The implementation of the correction terms and low-order equation is set up to allow the reuse of code for WLS and \saaft implementations.\par
    
    Given a specific mesh and an angular quadrature, the left hand side operator of both WLS and \saaft equation are fixed, and therefore the matrix can be assembled during the initial setup. Since the low order and high-order system use the same mesh, the spatial quadrature points are identical. All coefficients and drift vectors are only evaluated on the quadrature points and then transfered to the low-order system. The low-order system uses the same code for both high-order schemes. Code duplications were also avoided for the boundary coefficient and transfer routines, only the evaluation of the drift vector is different between WLS and SAAF.\par
    
    All one-dimensional results were generated using a simpler Python research code. This code allowed better control and easier modifications. The numerical solver was the SciPy sparse linear algebra solver. \par

\subsection{Comparison of weighted and unweighted LS}

    To test the effect of the weighting on the LS equation, consider a one dimensional problem with two material regions. The left region contains a weak absorber (\(\sigt[1] = \siga[1] = 0.1\Xsunit\)), while the right region has a strong absorber (\(\sigt[2] = \siga[2] = 10\Xsunit\)). Each region is 1\cm thick and discretized with 8 cell. The problem is surrounded by vacuum. A constant source of \(q = 1\frac{\unit{n}}{\unit{s}\cdot\unit{cm}}\) is added in both regions. We compared the unweighted LS to the weighted LS and the SAAF and \saaft formulation in \rattlesnake. The \saaft scheme used \(\zeta = 0.5\), and all calculations employed a \sn[8] Gauss quadrature. Note that \weight[\mathrm{max}] is not needed, since the problem does not contain a void. \par
    
    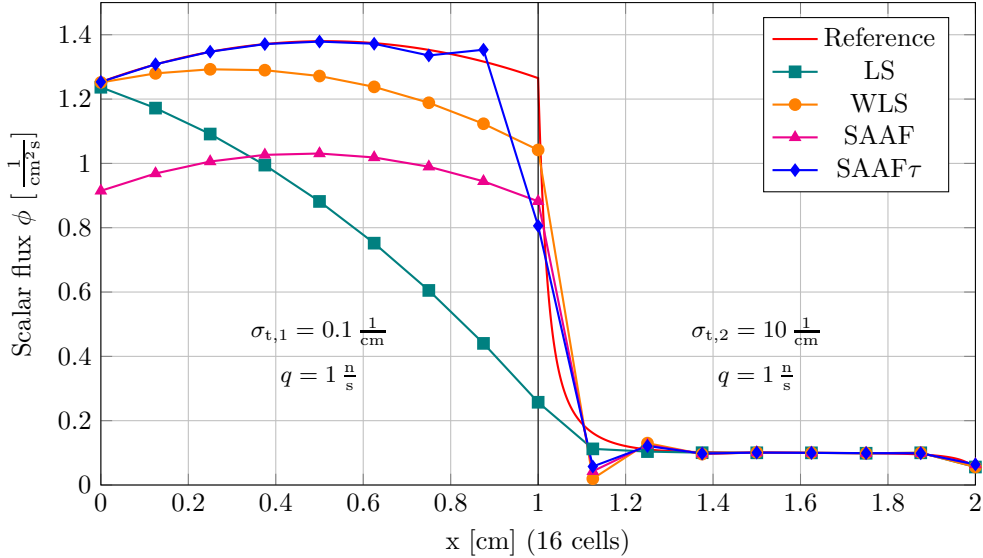
\begin{figure}[tp]
    	\setlength{\figheight}{8cm}
    	\begin{minipage}{0.8\textwidth}
    		\begin{tikzpicture}
  \begin{axis}[
        xmin = 0.0, xmax = 2,
        ymin = 0.0, ymax = 1.5,
        height=\figheight,
        width=\textwidth,
        grid=major,
        legend pos = north east,
        legend columns = 1,
        xlabel = {x [cm] (16 cells)},
        ylabel = {Scalar flux \(\phi\) [\sfluxunit]}
    ]

    \addplot [red, thick] table [x = x, y = phi, col sep=comma] {data/fig_1_2/two_region_source_reference.csv};
    \addlegendentry{Reference}

    \addplot [teal, solid, thick, mark = square*] table [x = x, y = phi, col sep=comma] {data/fig_1_2/two_region_source_ls.csv};
    \addlegendentry{LS}

    \addplot [orange, solid, thick, mark = *] table [x = x, y = phi, col sep=comma] {data/fig_1_2/two_region_source_wls.csv};
    \addlegendentry{WLS}

    \addplot [magenta, thick, mark = triangle*] table [x = x, y = phi, col sep=comma] {data/fig_1_2/two_region_source_saaf.csv};
    \addlegendentry{SAAF}

    \addplot [blue, solid, thick, mark = diamond*] table [x = x, y = phi, col sep=comma] {data/fig_1_2/two_region_source_saaft.csv};
    \addlegendentry{\saaft}

    \addplot [black, thin, below] coordinates {(1, 0) (1, 1.5)};

	\pgfplotsset{
        after end axis/.code={
            \node[black, below] at (axis cs:0.5,0.55){\small{\(\sigt[1] = 0.1\Xsunit\)}};
            \node[black, below] at (axis cs:0.5,0.4){\small{\(q = 1\sourceunit\)}};

            \node[black, below] at (axis cs:1.5,0.55){\small{\(\sigt[2] = 10\Xsunit\)}};
            \node[black, below] at (axis cs:1.5,0.4){\small{\(q = 1\sourceunit\)}};
        }
    }
  \end{axis}
\end{tikzpicture}
    	\end{minipage}
    	\caption{Comparison of the scalar flux results for the two absorber problem with a source using the second order transport schemes.}
    	\label{fig:constant_weight_angular_flux}
    \end{figure}
    
    The purpose of these test problems were to evaluate the pure LS and WLS equations. No NDA was used, which would enforce conservation, even though acceleration is unnecessary in a purely absorbing geometry. \Cref{fig:constant_weight_angular_flux} shows the results for the scalar flux. The LS result in the left half of the problem was strongly influenced by the thick material in the right half. The introduction of the weight function for the WLS ameliorated this problem. The results still show a decrease towards the thick region, however it is significantly less compared to the unweighted LS. The reason, that the WLS scheme gives a different results than the SAAF is the boundary condition, \cref{eq:ls_bc_7}. The cell size is too large, so the void guard is triggered. If we use \cref{eq:ls_bc_1} instead, we obtain the same result as the SAAF calculations. The \saaft scheme was closest to the reference solution, but it had a strong decrease in the cell next to the material interface and oscillations left of that cell. We note the large difference between the SAAF method and the \saaft method, even though no voids are present. With the default setting \(\zeta = 0.5\), the void stabilization is already activated for most reactor physics problems and hence influences the results. \par
    
        \begin{figure}[t]
        \setlength{\figheight}{8cm}
        \begin{minipage}{0.8\textwidth}
                
\begin{tikzpicture}
  \begin{axis}[
        xmin = 0.0, xmax = 2,
        ymin = -0.5, ymax = 2.5,
        height=\figheight,
        width=\textwidth,
        grid=major,
    	legend pos = north east,
        legend columns = 2,
        cycle list name = color list,
        xlabel = {x [cm] (16 cells)},
        ylabel = {Angular flux \(\psi\) [\afluxunit]},
        mark options={solid, scale = 0.75}
    ]

    \addplot [thick, red, dashed, mark = square*, mark repeat=128]table [x = x, y = mu_5, col sep=comma] {data/fig_1_2/two_region_source_reference.csv};
    \addlegendentry{\(\mu_5\) ref}

    \addplot [thick, red, mark = square*] table [x = x, y = mu_5, col sep=comma] {data/fig_1_2/two_region_source_wls.csv};
    \addlegendentry{\(\mu_5\) WLS}

    \addplot [thick, blue, dashed, mark = *, mark repeat=128] table [x = x, y = mu_6, col sep=comma] {data/fig_1_2/two_region_source_reference.csv};
    \addlegendentry{\(\mu_6\) ref}

    \addplot [thick, blue, mark = *] table [x = x, y = mu_6, col sep=comma] {data/fig_1_2/two_region_source_wls.csv};
    \addlegendentry{\(\mu_6\) WLS}

    \addplot [thick, teal, dashed, mark = triangle*, mark repeat=128] table [x = x, y = mu_7, col sep=comma] {data/fig_1_2/two_region_source_reference.csv};
    \addlegendentry{\(\mu_7\) ref}

    \addplot [thick, teal, mark = triangle*] table [x = x, y = mu_7, col sep=comma] {data/fig_1_2/two_region_source_wls.csv};
    \addlegendentry{\(\mu_7\) WLS}

    \addplot [thick, orange, dashed, mark = diamond*, mark repeat=128] table [x = x, y = mu_8, col sep=comma] {data/fig_1_2/two_region_source_reference.csv};
    \addlegendentry{\(\mu_8\) ref}

    \addplot [thick, orange, mark = diamond*] table [x = x, y = mu_8, col sep=comma] {data/fig_1_2/two_region_source_wls.csv};
    \addlegendentry{\(\mu_8\) WLS}

    \addplot [black, thin, below] coordinates {(1, -1) (1, 3)};

    \pgfplotsset{
        after end axis/.code={
            \node[black, below] at (axis cs:0.5,2.5){\small{\(\sigt[1] = 0.1\Xsunit\)}};
            \node[black, below] at (axis cs:0.5,2.25){\small{\(q = 1\sourceunit\)}};

            \node[black, below] at (axis cs:1.5,1.05){\small{\(\sigt[2] = 10\Xsunit\)}};
            \node[black, below] at (axis cs:1.5,0.75){\small{\(q = 1\sourceunit\)}};
        }
    }

  \end{axis}
\end{tikzpicture}
        \end{minipage}
        \caption{Comparison of the angular flux for  positive angles between the WLS (solid line) and the reference solution (dashed) depending on the angle \(\mu\) (\(0 < \mu_5 < \mu_6 < \mu_7 < \mu_8 < 1\)).}
        \label{fig:aflux_weight}
    \end{figure}
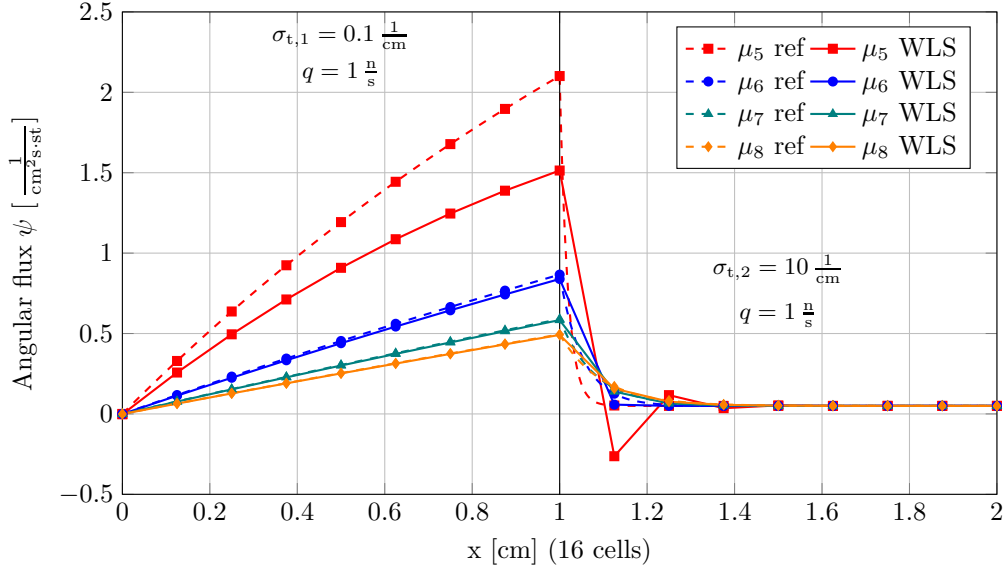
    
    The results for the angular fluxes in \cref{fig:aflux_weight} showed that the error for the WLS scheme is strongly dependent on the angle \(\mu\). For more perpendicular directions the error was larger (\cref{fig:aflux_weight}) than for \(\mu\) closer to one. Note that the angular flux can be negative, because the WLS scheme is not strictly non-negative since the matrix is not monotone.
    
    \subsection{Weight function limit in voids}
    
    \begin{figure}[tp]
    	\setlength{\figheight}{8cm}
    	\begin{minipage}{0.8\textwidth}
    		\begin{tikzpicture}
  \begin{axis}[
        xmin = 0.0, xmax = 2,
        ymin = -0.1, ymax = 0.6,
        height=\figheight,
        width=\textwidth,
        grid=major,
    	legend pos = north east,
        legend columns = 1,
        cycle list name = color list,
        xlabel = {x [cm] 16 cells},
        ylabel = {Scalar flux \(\phi\) [\sfluxunit]},
        mark options={solid, scale = 0.75}
    ]

    \addplot [thick, red, solid]table [x = x, y = analytic, col sep=comma] {data/fig_3.csv};
    \addlegendentry{Analytic}

    \addplot [thick, blue, solid, mark = square*]table [x = x, y = w_1, col sep=comma] {data/fig_3.csv};
    \addlegendentry{\(\weight[\mathrm{max}] = 1\cm\)}

    \addplot [thick, orange, solid, mark = *]table [x = x, y = w_5, col sep=comma] {data/fig_3.csv};
    \addlegendentry{\(\weight[\mathrm{max}] = 5\cm\)}

    \addplot [thick, magenta, solid, mark = triangle*]table [x = x, y = w_10, col sep=comma] {data/fig_3.csv};
    \addlegendentry{\(\weight[\mathrm{max}] = 10\cm\)}

    \addplot [thick, teal, solid, mark = +]table [x = x, y = w_50, col sep=comma] {data/fig_3.csv};
    \addlegendentry{\(\weight[\mathrm{max}] = 50\cm\)}

    \addplot [black, thin, below] coordinates {(1, -1) (1, 1)};

    \pgfplotsset{
        after end axis/.code={
            \node[black, below] at (axis cs:0.24,0.6){\small{\(\phi^\mathrm{inc} = 1.0\sfluxunit\)}};
            \node[black, below] at (axis cs:0.5,0.18){\small{\(\sigt[1] = 0.0\Xsunit\)}};
            \node[black, below] at (axis cs:1.5,0.18){\small{\(\sigt[2] = 10\Xsunit\)}};
        }
    }

  \end{axis}
\end{tikzpicture}
    	\end{minipage}
    	\caption{Scalar fluxes using the WLS transport scheme for different weight function limit \weight[\mathrm{max}] for the two region void problem with left inflow.}
    	\label{fig:wls_max_weight}
    \end{figure}
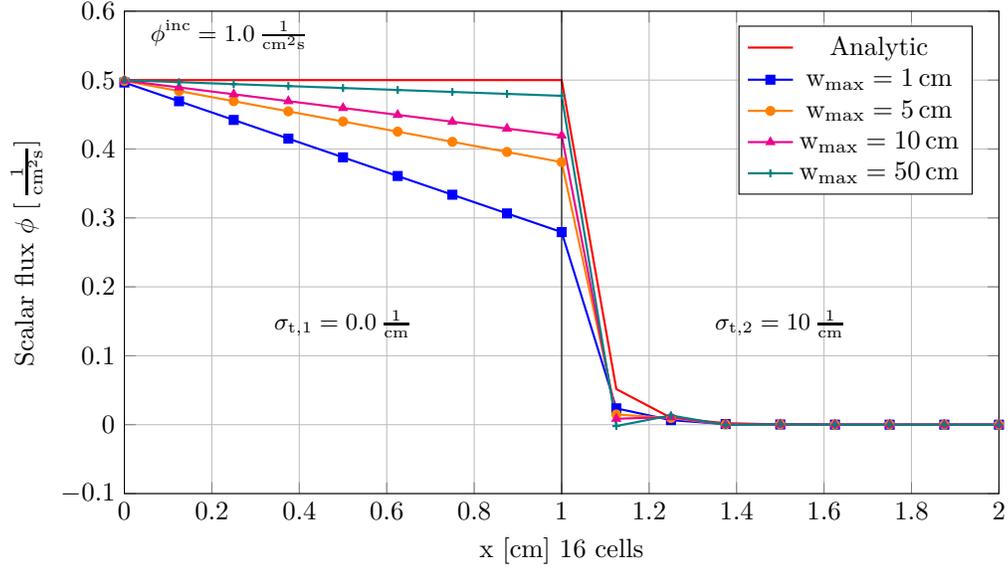
    		
    \begin{figure}[tp]
    	\setlength{\figheight}{8cm}
    	\begin{minipage}{0.8\textwidth}
    		\begin{tikzpicture}
    \begin{loglogaxis}[
       		xmin = 1, xmax = 1e6,
       		ymin = 1e-6, ymax = 1,
            ytick = {1e-6,1e-5,1e-4,1e-3,1e-2,1e-1,1},
            y tick label style={/pgf/number format/.cd, scaled y ticks = false, fixed},
            x tick label style={
                xticklabel={
                    \pgfkeys{/pgf/fpu=true}
                    \pgfmathparse{exp(\tick)}
                    \pgfmathprintnumber[fixed relative, precision=3]{\pgfmathresult}
                    \pgfkeys{/pgf/fpu=false}
                }
            },
            height=\figheight,
            width=\textwidth,
            grid=major,
            legend pos = south west,
            legend columns = 1,
            cycle list name = exotic,
            xlabel = {Maximum weight function \(\weight_\mathrm{max}\) [\cm]},
            ylabel = {Relative \(\mathrm{L}_2\) error [-]},
            mark options={solid, scale = 0.75}
        ]

        \addplot [red, thick, mark = square*] table [x = x, y = complete, col sep=comma] {data/fig_4.csv};
        \addlegendentry{left, total}

        \addplot [blue, dashed, thick, mark = triangle] table [x = x, y = vacuum, col sep=comma] {data/fig_4.csv};
        \addlegendentry{left, vacuum}

        \addplot [orange, solid, thick, mark = *] table [x = x, y = complete, col sep=comma] {data/fig_4_1.csv};
        \addlegendentry{right, total}

        \addplot [teal, dashed, thick, mark = o] table [x = x, y = vacuum, col sep=comma] {data/fig_4_1.csv};
        \addlegendentry{right, vacuum}

    \end{loglogaxis}
\end{tikzpicture}
    	\end{minipage}
    	\caption{Error convergence with increasing weight function limit \weight[\mathrm{max}] for the two region void problem with left or right inflow for the total error and the error in the vacuum region separately.}
    	\label{fig:wls_max_weight_convergence}
    \end{figure}
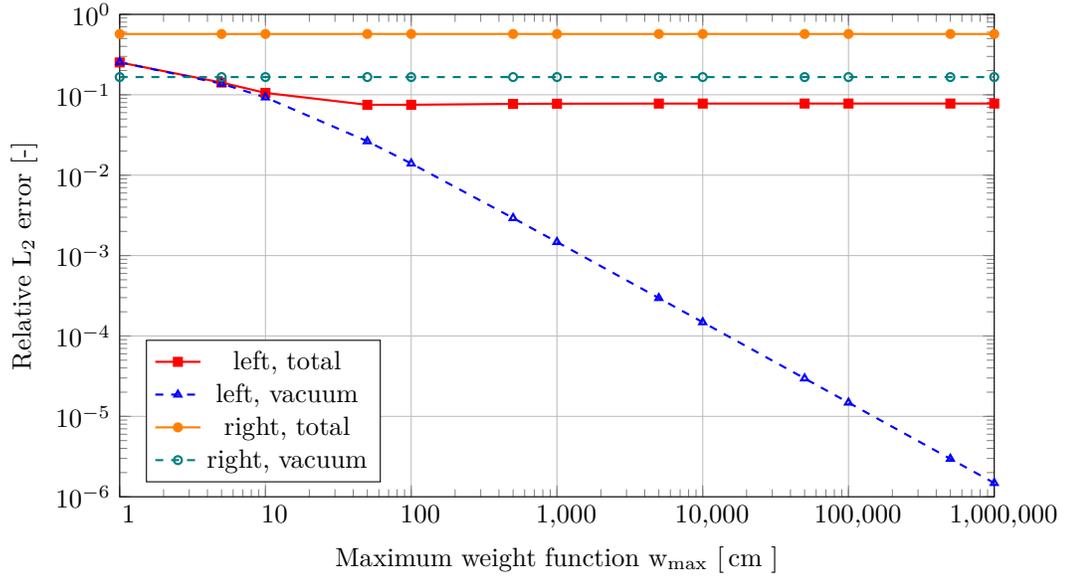
    				
    As stated in \cref{sec:wls}, the weight function must be limited to be defined in void regions. We further studied the influence of the maximal weight \weight[\mathrm{max}] of the WLS implementation in a problem with void to obtain an estimate of \weight[\mathrm{max}] necessary for good accuracy. If \weight[\mathrm{max}] is too small, the accuracy will be low, however, the larger \weight[\mathrm{max}] gets, the more the discretization matrix becomes ill-conditioned.
    
    The test problem is a two region slab with a void (\(\sigt[1] = 0\Xsunit\)) on the left side and a strong absorber (\(\sigt[2] = 10\Xsunit\)) on the right side. Two subcases with an isotropic flux of \(\phi^\mathrm{inc} = 1\sfluxunit\) on the left or the right boundary respectively demonstrate the directionality of the problem. For all calculations a \sn[8] quadrature were used. \par
    
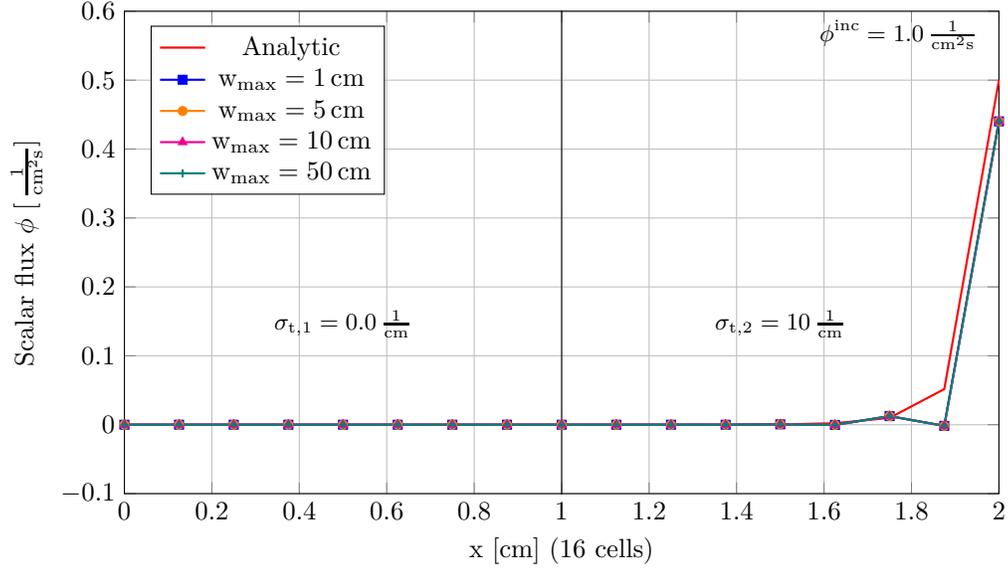
\begin{figure}[tp]
        \setlength{\figheight}{8cm}
        \begin{minipage}{0.8\textwidth}
            \begin{tikzpicture}
  \begin{axis}[
        xmin = 0.0, xmax = 2,
        ymin = -0.1, ymax = 0.6,
        height=\figheight,
        width=\textwidth,
        grid=major,
    	legend pos = north west,
        legend columns = 1,
        cycle list name = color list,
        xlabel = {x [cm] (16 cells)},
        ylabel = {Scalar flux \(\phi\) [\sfluxunit]},
        mark options={solid, scale = 0.75}
    ]

    \addplot [thick, red, solid]table [x = x, y = analytic, col sep=comma] {data/fig_5.csv};
    \addlegendentry{Analytic}

    \addplot [thick, blue, solid, mark = square*]table [x = x, y = w_1, col sep=comma] {data/fig_5.csv};
    \addlegendentry{\(\weight[\mathrm{max}] = 1\cm\)}

    \addplot [thick, orange, solid, mark = *]table [x = x, y = w_5, col sep=comma] {data/fig_5.csv};
    \addlegendentry{\(\weight[\mathrm{max}] = 5\cm\)}

    \addplot [thick, magenta, solid, mark = triangle*]table [x = x, y = w_10, col sep=comma] {data/fig_5.csv};
    \addlegendentry{\(\weight[\mathrm{max}] = 10\cm\)}

    \addplot [thick, teal, solid, mark = +]table [x = x, y = w_50, col sep=comma] {data/fig_5.csv};
    \addlegendentry{\(\weight[\mathrm{max}] = 50\cm\)}

    \addplot [black, thin, below] coordinates {(1, -1) (1, 1)};

    \pgfplotsset{
        after end axis/.code={
            \node[black, below] at (axis cs:1.77,0.6){\small{\(\phi^\mathrm{inc} = 1.0\sfluxunit\)}};
            \node[black, below] at (axis cs:0.5,0.18){\small{\(\sigt[1] = 0.0\Xsunit\)}};
            \node[black, below] at (axis cs:1.5,0.18){\small{\(\sigt[2] = 10\Xsunit\)}};
        }
    }

  \end{axis}
\end{tikzpicture}
        \end{minipage}
        \caption{WLS scalar fluxes results for different weight function limits \weight[\mathrm{max}] for the two region void problem with right inflow.}
        \label{fig:wls_max_weight_reversed}
    \end{figure}

    \Cref{fig:wls_max_weight} shows the scalar flux results for the left inflow case. The increase of the maximal weight improved the slope of the scalar flux significantly in the void region. However, the increase resulted in a stronger dip after the material interface. The reduction in the relative error
    \begin{equation} \label{eq:def_relative_error}
    	e \equiv \frac{\norm[L_2]{\phi\left(x\right) - \phi_\mathrm{exact}\left(x\right)}}{\norm[L_2]{\phi_\mathrm{exact}\left(x\right)}}
    \end{equation}
    with increasing \weight[\mathrm{max}] is shown in \cref{fig:wls_max_weight_convergence}. The reference solution is a spatially analytic solution using the same angular quadrature. The convergence is shown for the error of the whole domain (void and material region) and for the void region separately. The ratio between these two errors is a good indicator for the effectiveness of the weighting, since \weight[\mathrm{max}] strongly influences the void region, but has almost no effect in the material region except for the first few nodes. We can see that for the left inflow case the error in the void converged with first order for increasing \weight[\mathrm{max}]. For \(\weight[\mathrm{max}] < 10\cm \) the error in the void region dominated the error for the whole problem, however for larger \weight[\mathrm{max}] the error in the material region dominated and hence no further error reduction can be seen. From this we concluded that a \weight[\mathrm{max}] in the range between 100\cm and 1000\cm is sufficient for this problem. For the right inflow problem \cref{fig:wls_max_weight_reversed} shows that \weight[\mathrm{max}] had no effect on the error in both the whole domain and void region. This clearly demonstrates the directionality of the causality problem.
    
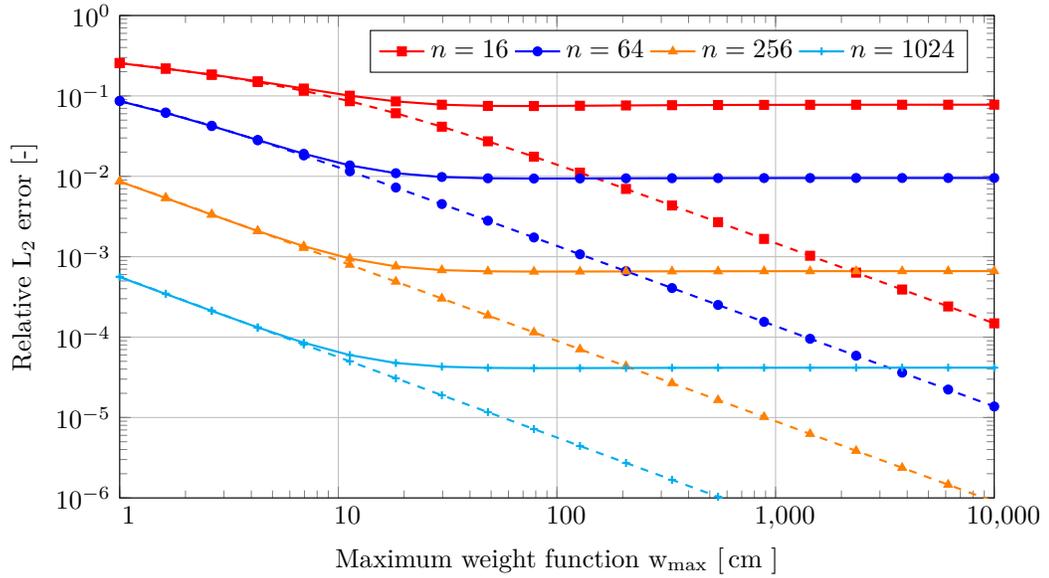
\begin{figure}[tp]
    	\setlength{\figheight}{8cm}
    	\begin{minipage}{0.8\textwidth}
    		\begin{tikzpicture}
    \begin{loglogaxis}[
       		xmin = 1, xmax = 1e4,
       		ymin = 1e-6, ymax = 1,
            ytick = {1e-6,1e-5,1e-4,1e-3,1e-2,1e-1,1},
            y tick label style={/pgf/number format/.cd, scaled y ticks = false, fixed},
            x tick label style={
                xticklabel={
                    \pgfkeys{/pgf/fpu=true}
                    \pgfmathparse{exp(\tick)}
                    \pgfmathprintnumber[fixed relative, precision=3]{\pgfmathresult}
                    \pgfkeys{/pgf/fpu=false}
                }
            },
            height=\figheight,
            width=\textwidth,
            grid=major,
            legend pos = north east,
            legend columns = 4,
            cycle list name = exotic,
            xlabel = {Maximum weight function \(\weight_\mathrm{max}\) [\cm]},
            ylabel = {Relative \(\mathrm{L}_2\) error [-]},
            mark options={solid, scale = 0.75}
        ]

        \addplot [red, thick, mark = square*] table [x = w_max, y = n_16, col sep=comma] {data/fig_6.csv};
        \addlegendentry{\(n = 16\)}
        
        \addplot [red, thick, dashed, mark = square*, forget plot] table [x = w_max, y = n_16_vac, col sep=comma] {data/fig_6.csv};
        
        \addplot [blue, thick, mark = *] table [x = w_max, y = n_64, col sep=comma] {data/fig_6.csv};
        \addlegendentry{\(n = 64\)}
        
        \addplot [blue, thick, dashed, mark = *, forget plot] table [x = w_max, y = n_64_vac, col sep=comma] {data/fig_6.csv};
        
        \addplot [orange, thick, mark = triangle*] table [x = w_max, y = n_256, col sep=comma] {data/fig_6.csv};
        \addlegendentry{\(n = 256\)}
        
        \addplot [orange, thick, dashed, mark = triangle*, forget plot] table [x = w_max, y = n_256_vac, col sep=comma] {data/fig_6.csv};
        
        \addplot [cyan, thick, mark = +] table [x = w_max, y = n_1024, col sep=comma] {data/fig_6.csv};
        \addlegendentry{\(n = 1024\)}
        
        \addplot [cyan, thick, dashed, mark = +, forget plot] table [x = w_max, y = n_1024_vac, col sep=comma] {data/fig_6.csv};

    \end{loglogaxis}
\end{tikzpicture}
    	\end{minipage}
    	\caption{The error convergence with increasing weight function limit \weight[\mathrm{max}] using different mesh sizes \(n\) for the two region void problem with left inflow (solid line for the total error, dashed for the error in the vacuum region).}
    	\label{fig:wls_max_weight_convergence_h}
    \end{figure}

    The values for \weight[\mathrm{max}] are problem dependent. In the following section, we evaluate the influence of several parameters on the required \weight[\mathrm{max}]. Again we will look at the ratio of the error in the void region to the total error as a function of \weight[\mathrm{max}].  The first parameter we looked at is the mesh size. We ran a study on the effects of mesh refinement on the optimal limit of the weight function. Our results showed that the optimal weight limit is almost independent from the mesh refinement as shown in \cref{fig:wls_max_weight_convergence_h}. In the figure, the solid lines show the error for the whole problem, the dashed lines the error for the vacuum part of the problem. For all mesh sizes we observed the same behavior, that at approximately \(\weight[\mathrm{max}] = 100\cm\) the error in the void is almost a magnitude lower that the total error and hence its contribution to the total error is negligible and other error sources dominate. This indicates that the mesh size has no significant effect on the required \weight[\mathrm{max}].\par

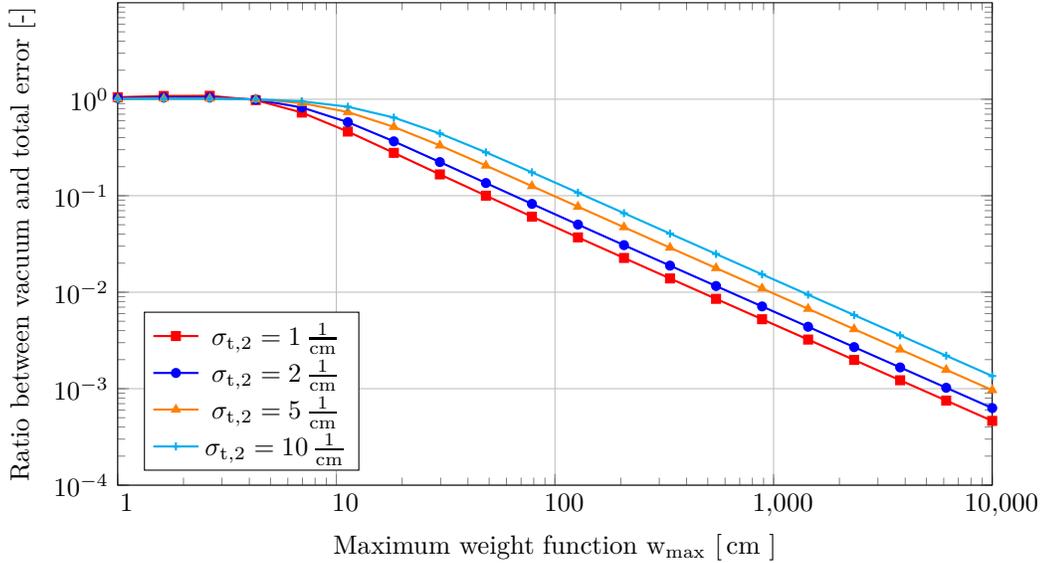
\begin{figure}[tp]
        \setlength{\figheight}{8cm}
        \begin{minipage}{0.8\textwidth}
            \begin{tikzpicture}
    \begin{loglogaxis}[
       		xmin = 1, xmax = 1e4,
       		ymin = 1e-4, ymax = 1e1,
            ytick = {1e-10, 1e-9,1e-8,1e-7,1e-6,1e-5,1e-4,1e-3,1e-2,1e-1, 1e0},
            y tick label style={/pgf/number format/.cd, scaled y ticks = false, fixed},
            x tick label style={
                xticklabel={
                    \pgfkeys{/pgf/fpu=true}
                    \pgfmathparse{exp(\tick)}
                    \pgfmathprintnumber[fixed relative, precision=3]{\pgfmathresult}
                    \pgfkeys{/pgf/fpu=false}
                }
            },
            height=\figheight,
            width=\textwidth,
            grid=major,
            legend pos = south west,
            legend columns = 1,
            cycle list name = exotic,
            xlabel = {Maximum weight function \(\weight_\mathrm{max}\) [\cm]},
            ylabel = {Ratio between vacuum and total error [-]},
            mark options={solid, scale = 0.75}
        ]

        \addplot [red, thick, mark = square*] table [x = w_max, y = xs_1.0_ratio, col sep=comma] {data/fig_9.csv};
        \addlegendentry{\(\sigt[2] = 1\Xsunit\)}
        
        \addplot [blue, thick, mark = *] table [x = w_max, y = xs_2.0_ratio, col sep=comma] {data/fig_9.csv};
        \addlegendentry{\(\sigt[2] = 2\Xsunit\)}
        
        \addplot [orange, thick, mark = triangle*] table [x = w_max, y = xs_5.0_ratio, col sep=comma] {data/fig_9.csv};
        \addlegendentry{\(\sigt[2] = 5\Xsunit\)}
        
        \addplot [cyan, thick, mark = +] table [x = w_max, y = xs_10.0_ratio, col sep=comma] {data/fig_9.csv};
        \addlegendentry{\(\sigt[2] = 10\Xsunit\)}

    \end{loglogaxis}
\end{tikzpicture}
        \end{minipage}
        \caption{Ratio of the error in the vacuum region to the total error as a function of \weight[max] with different \(\sigt[2]\) of the material region for the two region problem with \(n = 1024\), \(L_1 = 1\cm\) and \(L_2 = 1\cm\)}
        \label{fig:wls_max_weight_convergence_xs}
    \end{figure}

    The second set of calculations addresses the influence of geometric parameters on the optimal \weight[\mathrm{max}]. The base problem used for these calculations is similar to the previous problem. It is a two region problem with void on the left (\(\sigt[1] = \siga[1] = 0\Xsunit, L_1 = 1\cm\)) and an absorber on the right with \(\sigt[2] = \siga[2] = 1\Xsunit\) and a width of \(L_2 = 1\cm\). By changing each of the parameters \(L_1\), \(L_2\) and \sigt[2] separately and locking at the ratio of the error in the void region to the total error we can study their sensitivity on the required maximal weight. Every calculation used 1024 cells to provide a good spatial resolution. \Cref{fig:wls_max_weight_convergence_xs,fig:wls_max_weight_convergence_left,fig:wls_max_weight_convergence_right} show the results for this parameter study. The red line in every plot is the default configuration without changed parameters as described above. A ratio close to one shows that the error in the void region is dominant, while a small ratio indicates a small influence of the void error and that other sources of error dominate. \par
    
    \begin{figure}[tp]
        \setlength{\figheight}{8cm}
        \begin{minipage}{0.8\textwidth}
            \begin{tikzpicture}
    \begin{loglogaxis}[
       		xmin = 1, xmax = 1e4,
               ymin = 1e-4, ymax = 1e1,
               ytick = {1e-10, 1e-9,1e-8,1e-7,1e-6,1e-5,1e-4,1e-3,1e-2,1e-1, 1e0},
            y tick label style={/pgf/number format/.cd, scaled y ticks = false, fixed},
            x tick label style={
                xticklabel={
                    \pgfkeys{/pgf/fpu=true}
                    \pgfmathparse{exp(\tick)}
                    \pgfmathprintnumber[fixed relative, precision=3]{\pgfmathresult}
                    \pgfkeys{/pgf/fpu=false}
                }
            },
            height=\figheight,
            width=\textwidth,
            grid=major,
            legend pos = south west,
            legend columns = 1,
            cycle list name = exotic,
            xlabel = {Maximum weight function \(\weight_\mathrm{max}\) [\cm]},
            ylabel = {Ratio between vacuum and total error [-]},
            mark options={solid, scale = 0.75}
        ]

        \addplot [red, thick, mark = square*] table [x = w_max, y = l_1_ratio, col sep=comma] {data/fig_7.csv};
        \addlegendentry{\(L_1 = 1\cm\)}
        
        \addplot [blue, thick, mark = *] table [x = w_max, y = l_2_ratio, col sep=comma] {data/fig_7.csv};
        \addlegendentry{\(L_1 = 2\cm\)}
        
        \addplot [orange, thick, mark = triangle*] table [x = w_max, y = l_5_ratio, col sep=comma] {data/fig_7.csv};
        \addlegendentry{\(L_1 = 5\cm\)}
        
        \addplot [cyan, thick, mark = +] table [x = w_max, y = l_10_ratio, col sep=comma] {data/fig_7.csv};
        \addlegendentry{\(L_1 = 10\cm\)}

    \end{loglogaxis}
\end{tikzpicture}
        \end{minipage}
        \caption{Ratio of the error in the vacuum region to the total error as a function of \weight[max] with different widths \(L_1\) of the vacuum region for the two region problem with \(n = 1024\), \(\sigt[2] = 1\Xsunit\) and \(L_2 = 1\cm\)}
        \label{fig:wls_max_weight_convergence_left}
    \end{figure}
    
    \begin{figure}[tp]
        \setlength{\figheight}{8cm}
        \begin{minipage}{0.8\textwidth}
            \begin{tikzpicture}
    \begin{loglogaxis}[
       		xmin = 1, xmax = 1e4,
            ymin = 1e-4, ymax = 1e1,
            ytick = {1e-10, 1e-9,1e-8,1e-7,1e-6,1e-5,1e-4,1e-3,1e-2,1e-1, 1e0},
            y tick label style={/pgf/number format/.cd, scaled y ticks = false, fixed},
            x tick label style={
                xticklabel={
                    \pgfkeys{/pgf/fpu=true}
                    \pgfmathparse{exp(\tick)}
                    \pgfmathprintnumber[fixed relative, precision=3]{\pgfmathresult}
                    \pgfkeys{/pgf/fpu=false}
                }
            },
            height=\figheight,
            width=\textwidth,
            grid=major,
            legend pos = south west,
            legend columns = 1,
            cycle list name = exotic,
            xlabel = {Maximum weight function \(\weight_\mathrm{max}\) [\cm]},
            ylabel = {Ratio between vacuum and total error [-]},
            mark options={solid, scale = 0.75}
        ]

        \addplot [red, thick, mark = square*] table [x = w_max, y = l_1_ratio, col sep=comma] {data/fig_8.csv};
        \addlegendentry{\(L_2 = 1\cm\)}
        
        \addplot [blue, thick, mark = *] table [x = w_max, y = l_2_ratio, col sep=comma] {data/fig_8.csv};
        \addlegendentry{\(L_2 = 2\cm\)}
        
        \addplot [orange, thick, mark = triangle*] table [x = w_max, y = l_5_ratio, col sep=comma] {data/fig_8.csv};
        \addlegendentry{\(L_2 = 5\cm\)}
        
        \addplot [cyan, thick, mark = +] table [x = w_max, y = l_10_ratio, col sep=comma] {data/fig_8.csv};
        \addlegendentry{\(L_2 = 10\cm\)}

    \end{loglogaxis}
\end{tikzpicture}
        \end{minipage}
        \caption{Ratio of the error in the vacuum region to the total error as a function of \weight[max] with different widths \(L_2\) of the material region for the two region problem with \(n = 1024\), \(\sigt[2] = 1\Xsunit\) and \(L_1 = 1\cm\)}
        \label{fig:wls_max_weight_convergence_right}
    \end{figure} \par
    
    \Cref{fig:wls_max_weight_convergence_left} shows that the width of the vacuum region has the strongest influence on the needed \weight[\mathrm{max}], while the width of the material region and the corresponding cross section have both a smaller influence (\cref{fig:wls_max_weight_convergence_right,fig:wls_max_weight_convergence_xs}). This can be easier seen in \cref{tab:wls_max_intersect}, which shows the \weight[\mathrm{max}] for which the error in the vacuum regions is only 0.1 of the total error. While an increase in \(L_1\) of a factor of ten results in an increase of the required weight limit of the same factor, this factor is only approximately 2.8 for both \(L_2\) and \sigt[2].
    
    \begin{table}[tp]
        \caption{\weight[\mathrm{max}] for which the error in the vacuum is 10\% of the total error based on the parameters \(L_1\), \(L_2\) and \sigt[2].}
        \label{tab:wls_max_intersect}
        \begin{tabular}{lrrr}
        	\toprule
        	\multicolumn{1}{c}{Value of parameter} & \multicolumn{1}{c}{\(\weight[\mathrm{max}]\left(L_1\right)\)} & \multicolumn{1}{c}{\(\weight[\mathrm{max}]\left(L_2\right)\)} & \multicolumn{1}{c}{\(\weight[\mathrm{max}]\left(\sigt[2]\right)\)} \\
            \multicolumn{1}{c}{\(L_1\)[\cm], \(L_2\)[\cm], \sigt[2] [\(\cm^{-1}\)]} & \multicolumn{1}{c}{[\cm]} & \multicolumn{1}{c}{[\cm]} & \multicolumn{1}{c}{[\cm]} \\
             \midrule
        	1                         &                        48.4 &                        48.4 &                             48.4 \\
        	2                         &                        96.8 &                        64.9 &                             64.9 \\
        	5                         &                       242.1 &                        98.4 &                             98.4 \\
        	10                        &                       484.1 &                       136.8 &                            136.8 \\ \bottomrule
        \end{tabular}
    \end{table}
    
    We showed already, that the error of the angular flux is strongly dependent on the angle (\cref{fig:aflux_weight}). Given the fact, that the width of the void region is an important factor on how high the limit of the weight function must be and that the track length through the void is proportional to \(\mu^{-1}\), we can modify the limit to address the angular issue. Changing the weight function to
    \begin{equation} \label{eq:weight_fct_mu}
        \weight[m] \equiv \min\left(\frac{1}{\sigt}, \frac{\weight[max]}{\mu_m}\right)
    \end{equation}
    will change the \weight[\mathrm{max}] for a ratio of 10\% between vacuum error and total error to the values shown in \cref{tab:wls_max_intersect_mu}. This is a decrease by a factor of five compared to the previous values. This result is only meaningful in one-dimensional problems, the implementation in more dimensions is not clear. \par
    
    \begin{table}[tp]
        \caption{\weight[\mathrm{max}] with an angular dependent weight function (\cref{eq:weight_fct_mu}) for which the error in the vacuum is 10\% of the total error based on the parameters \(L_1\), \(L_2\) and \sigt[2].}
        \label{tab:wls_max_intersect_mu}
        \begin{tabular}{lrrr}
        	\toprule
            \multicolumn{1}{c}{Value of parameter} & \multicolumn{1}{c}{\(\weight[\mathrm{max}]\left(L_1\right)\)} & \multicolumn{1}{c}{\(\weight[\mathrm{max}]\left(L_2\right)\)} & \multicolumn{1}{c}{\(\weight[\mathrm{max}]\left(\sigt[2]\right)\)} \\
            \multicolumn{1}{c}{\(L_1\)[\cm], \(L_2\)[\cm], \sigt[2] [\(\cm^{-1}\)]} & \multicolumn{1}{c}{[\cm]} & \multicolumn{1}{c}{[\cm]} & \multicolumn{1}{c}{[\cm]} \\ \midrule
        	1                                      &                         9.7 &                         9.7 &                              9.7 \\
        	2                                      &                        19.4 &                        13.1 &                             13.1 \\
        	5                                      &                        48.6 &                        19.8 &                             19.8 \\
        	10                                     &                        97.2 &                        27.6 &                             27.6 \\ \bottomrule
        \end{tabular}
    \end{table}

\subsection{C5G7 benchmark}

    The C5G7 MOX benchmark problem is a challenging test for modern deterministic transport codes. We focused on the two dimensional version of the benchmark, which already requires large amounts of computational resources. The twenty sets of results that were initially submitted to the benchmark committee can be found in a special issue of Progress in Nuclear Energy~\cite{smith_benchmark_2004}. More recent calculations of the benchmark with a spatial and angular convergence study were presented by McGraw~\cite{mcgraw_accuracy_2015} using PDT and Wang~\cite{wang_convergence_2015} using \rattlesnake. These calculations were used as a reference solution to benchmark and validate the WLS implementation in Rattlesnake. \par
    
    The mesh is generated using the 2D mesh generator Triangle~\cite{shewchuk_triangle:_1996} with a geometry file which is created by a Rattlesnake mesh generator. The quality of the mesh ensures that no triangle has an interior angle less than 20 degrees. In order to limit the number of elements in the mesh, the surrounding reflector region is divided into three separate regions as shown in \cref{fig:c5g7_geomtry}, employing a coarser mesh far away from the fuel region, while the same maximum triangle area is applied to all fuel assemblies. The mesh conserves the volume of each fuel pin and hence the mass of fissile material. More details about the mesh can be found in the original paper~\cite{wang_convergence_2015}.\par

    \begin{figure}[tp]
        \caption{Zone layout of the C5G7 benchmark geometry. Reprinted from Wang et. al.~\cite{wang_convergence_2015}.}
        \label{fig:c5g7_geomtry}
    \end{figure}

	We performed calculations to compare the LS, WLS, SAAF and \saaft transport and NDA schemes. These calculations were performed with Gauss-Chebychev quadrature with 4 polar and 32 azimuthal angles. A angular refinement study showed that these settings were sufficient to minimize the angular error. For the transport solution a relative tolerance to the initial solution of \tento{-8} was used. The NDA calculations used as error tolerance the difference between two consecutive scalar flux solutions with a threshold of \tento{-8}, and a high order relative tolerance of \tento{-4}. These tolerances were used for all following calculations. \par
	
	\begin{table}[tp]
		\caption{Comparison of the eigenvalue error and the pin power errors for the transport and NDA schemes with 4 polar and 32 azimuthal angles. Column PP1 shows the relative error for the maximal pin power and PP2 the error for the minimal power.}
		\label{tab:c5g7_orignal_comparison}
		\begin{tabular}{lrrrrrrr}
			\toprule
			Scheme     & \(k_\mathrm{eff}\) &   AVG &   RMS &    MRE &    MAX &    PP1 &   PP2 \\
           &              [pcm] &  [\%] &  [\%] &   [\%] &   [\%] &   [\%] &  [\%] \\ \midrule
			LS         &          15880.916 & 8.354 & 0.336 & 11.640 & 30.847 & 12.345 & 4.255 \\
			WLS        &             87.741 & 0.400 & 0.016 &  0.539 &  1.661 &  0.608 & 1.123 \\
			SAAF       &             87.741 & 0.400 & 0.016 &  0.539 &  1.661 &  0.608 & 1.123 \\
			\saaft     &             52.115 & 0.249 & 0.010 &  0.350 &  1.178 &  0.433 & 0.195 \\ \midrule
			NDA LS     &             52.682 & 0.380 & 0.015 &  0.488 &  1.613 &  0.444 & 1.662 \\
			NDA WLS    &             87.740 & 0.400 & 0.016 &  0.539 &  1.661 &  0.608 & 1.123 \\
			NDA SAAF   &             87.741 & 0.400 & 0.016 &  0.539 &  1.661 &  0.608 & 1.123 \\
			NDA \saaft &             52.115 & 0.249 & 0.010 &  0.350 &  1.178 &  0.433 & 0.195 \\ \bottomrule
		\end{tabular}
	\end{table}
	
	The results are shown in \cref{tab:c5g7_orignal_comparison}. The non-conservative LS transport formulation showed large deviations in all errors. This demonstrates how important the use of a conservative scheme for criticality calculations is. The WLS transport solution was exactly the same as the SAAF solution, as expected for a problem without voids. The best result for all errors gave the \saaft scheme with \(\zeta = 0.5\), even though no voids are present in the benchmark. \par
	
	The results for the NDA schemes are shown in the second part of \cref{tab:c5g7_orignal_comparison}. The use of the conservative NDA with the non-conservative LS scheme showed that the NDA ensures conservation. The NDA LS result was within a reasonable range of error, with the error in the eigenvalue one of the lowest for all schemes. The errors in the pin power were also much lower compared to the LS transport solution, and of the same magnitude as the results of the NDA using SAAF and WLS. The NDA \saaft scheme gave the smallest errors for pin powers. Again SAAF and WLS, this time with NDA, gave the same results. The NDA results for WLS, SAAF and \saaft were consistent with the corresponding transport solves. \par
	
	\begin{table}[tp]
		\caption{Comparison of the pure transport and NDA calculation time and the number of NDA iterations for the original C5G7 benchmark.}
		\label{tab:c5g7_orignal_iterations}
		\begin{tabular}{l@{\hskip 40pt}r@{\hskip 40pt}rr}
			\toprule
			Scheme & \multicolumn{1}{c@{\hskip 40pt}}{Transport} &  \multicolumn{2}{c}{NDA}  \\
       &                                    Time [h] & Time [h] & Iterations [-] \\ \midrule
			LS     &                                        5.15 &     0.29 &             16 \\
			WLS    &                                        4.47 &     0.29 &             15 \\
			SAAF   &                                        4.25 &     0.28 &             15 \\
			\saaft &                                        4.79 &     0.23 &             15 \\ \bottomrule
		\end{tabular}
	\end{table}
	
	The comparison of the runtime for all schemes and the number of NDA iterations are shown in \cref{tab:c5g7_orignal_iterations}. All calculations were performed on 2 nodes of INL's HPC cluster Falcon~\cite{_idaho_2017} with 36 cores per node. The calculation times for pure transport were much longer than for the NDA schemes. \par
    
    \section{Conclusions}

    We derived a weighted LS transport equation and showed that we can make this equation equivalent to the SAAF equation by using a particular weight function. However, this weight function is not compatible with voids. So to be able to handle voids, the weight function and the boundary conditions were modified.  With these modifications, 
    the SAAF and weighted least-squares equations are equal only for sufficiently large cross sections. The primary 
    advantage of a weight function is an improvement in causality, i.e., a reduction in the extent to which downstream 
    quantities affect the transport solution.  The discretization of the WLS scheme is, in contrast to the \saaft scheme, symmetric positive definite in problems with voids. This, as mentioned earlier, allows the use of more memory efficient linear solvers with faster convergence. \par

    We analyzed limiting of the weight function to obtain best results for certain test problems.  Limiting is 
    required not only because the default weight function becomes unbounded in a void, but also because the 
    coefficient matrix becomes increasingly ill-conditioned as the variation of the weight function increases. Numerical evidence indicates that the weight function limit is a function of the geometry and strongly dependent on the width of the void region, while less impacted by the optical thickness of the surrounding regions. \par

    The weighted least-squares method and the corresponding NDA scheme were fully implemented in \rattlesnake. The C5G7 benchmark was used to test the NDA scheme on a more challenging 2-D problem. The comparison between NDA WLS, NDA \saaft, 
    and reference PDT calculations showed that the WLD NDA results are sometimes less accurate than the NDA \saaft results, but nonetheless comparable.  A major advantage of the NDA WLS scheme relative to the NDA \saaft scheme is that the NDA WLS high-order equations are symmetric positive-definite. Thus the NDA WLS scheme can use the conjugate-gradient method to solve the high-order equation, which requires the storage of only three solutions vectors.  More general Krylov solvers such as GMRES, can require an arbitrary number of solutions vectors or a restart with degraded convergence properties. This gives the NDA WLS scheme an enormous advantage regarding memory 
    requirements. \par

    Based on this work, we intend to continue testing the NDA WLS scheme on more complicated problems with and without voids. \par

    \section{Acknowledgments}
    This material is based upon work supported by the Department of Energy, Battelle Energy Alliance, LLC, under Award Number DE-AC07-05ID14517.
    
    \bibliographystyle{ans_js} 
    \bibliography{literature}
    
\end{document}